\documentclass[11pt]{article}
\renewcommand{\baselinestretch}{1.35}

\usepackage{amssymb,amsthm}

\newtheorem{definition}{Definition}
\newtheorem{lemma}{Lemma}
\newtheorem{proposition}{Proposition}
\newtheorem{cor}{Corollary}
\newtheorem{theorem}{Theorem}

\theoremstyle{definition}
 
\newtheorem{remark}{Remark}
\newtheorem{example}{Example}

\title{\bf Pseudo-potentials, nonlocal symmetries and integrability of some shallow water equations}
 
\author{Enrique G. Reyes}
 
\date{Department of Mathematics, University of Oklahoma\\      
      Norman, Oklahoma 73019 USA. \\
      E-Mail: ereyes@math.ou.edu}

\begin{document}
\maketitle
\begin{abstract}
Zero curvature formulations, pseudo-potentials, modified versions, ``Miura 
transformations'', and nonlocal symmetries of the Korteweg--de Vries, Camassa--Holm and 
Hunter--Saxton equations are investigated from an unified point of view: these three equations 
belong to a two--parameters family of equations ``describing pseudo-spherical surfaces'', and 
therefore their basic integrability properties can be studied by geometrical means. 
\end{abstract}

\section{Introduction}

The goal of this work is to present an unified account of some integrability properties 
of three important shallow water models, the Korteweg--de Vries, Camassa--Holm and Hunter--Saxton
equations. The main motivation behind this work comes from the papers \cite{bss,bss1,KM}: in 
\cite{bss,bss1} Beals, Sattinger and Szmigielski consider the scattering/inverse scattering 
analysis of these three equations from a unified perspective, and in \cite{KM} Khesin and 
Misio{\l}ek give an integrated account of their bi--hamiltonian formulations. In this article it
is pointed out that the existence of zero curvature formulations, quadratic pseudo--potentials, 
modified versions, ``Miura transformations'', and nonlocal symmetries for them, follow from some
developments linking differential geometry of surfaces and integrability of nonlinear partial 
differential equations \cite{ChT,R1,R1.5,T}.

One reason why these observations may be of importance (besides the fact that they show the 
usefulness of the geometric approach to integrability mentioned above) is 
that the construction of nonlocal symmetries carried out in this paper can be considered as a 
geometric implementation of the ``algebraic method'' used by M. Leo, R.A. Leo, G. Soliani, and 
P. Tempesta \cite{llst0,llst} to find nonlocal symmetries of nonlinear equations.

Recall that if $u_{t} = F$ is a scalar partial differential evolution equation in two independent
variables $x$ and $t$, a (generalized) symmetry of $u_{t} = F$ is a smooth 
function $G$ depending on $x,t,u$, and a finite number of derivatives of $u$ such that 
for any solution $u(x,t)$ of $u_{t}=F$, the deformed function $u(x,t) + \tau G(x,t)$ 
is also a solution to first order in $\tau$. At least at a formal level \cite[Chapter 5]{O} a 
generalized symmetry $G$ allows one to generate new solutions from old ones. If $G$ depends at 
most on $x,t,u,u_{x}$, one can indeed find a one--parameter group of transformations on the 
space of first order jets of the trivial bundle $(x,t,u) \mapsto (x,t)$  which ``sends solutions
to solutions'', see \cite{KVi,MSS,O}.

It appears that A. Vinogradov and I. Krasil'shchik \cite{VK80} were among the first to study
{\em nonlocal} symmetries of evolution equations rigorously, and to point out some of their
applications. By a nonlocal symmetry of $u_{t}=F$ one means (see Section 3 for a rigorous
definition) a function $G$ which depends on $x,t,u$, a finite number of $x$--derivatives of $u$ 
{\em and} (for example) indefinite integrals of $u$, such that for any solution $u(x,t)$ of
$u_{t} = F$, the function $u(x,t) + \tau G(u(x,t))$ is also a solution to first order in $\tau$. 
That these symmetries are both important and natural to consider has been increasingly 
acknowledged since Vinogradov and Krasil'shchik's paper \cite{VK80}. A few highlights are the
following: 

In 1982, Kaptsov \cite{Ka} solved the ``recursion operator equation'' $R_{t} = 
[F_{\ast},R]$, in which $F_{\ast}$ is the formal linearization of $F$ (see \cite{MSS,O} and 
Section 3 below) and found that his solution $R$ induced sequences of nonlocal symmetries of 
$u_{t}=F$. In the late 1980's, Bluman, Kumei and Reid \cite{BKR} and Bluman and Kumei \cite{BK} 
used nonlocal symmetries to find linearizing transformations for nonlinear equations. In 1991 
V.E. Adler \cite{Ad-r} introduced Lie algebras of nonlocal symmetries associated to equations 
integrable by the scattering/inverse scattering method, generalizing the local construction of 
integrable hierarchies due to M. Adler \cite{Ad}, Reimann and Semenov--Tyan--Shanski \cite{RS}, 
and others (see Faddeev and Takhtajan \cite{FT} for details and historical notes). Lastly, Galas 
\cite{Ga}, Leo et al. \cite{llst0,llst}, and Schiff \cite{s-pre} have quite recently obtained 
nonlocal symmetries of some well--known integrable equations, found their flows, and used them 
to construct special solutions for the equations at hand. 

An interesting characteristic of the papers by Galas, Leo {\em et. al.}, and Schiff, 
\cite{Ga,llst,s-pre}, is that the non-localities appearing in their symmetries are more involved 
than simply indefinite integrals of smooth functions of $x,t,u$ and a finite number of 
derivatives of $u$. In their examples, the nonlocal 
symmetries depend on {\em pseudo-potentials} of the equations they consider. These symmetries 
were anticipated and studied by Krasil'shchik and Vinogradov in the 1980's in the context of 
their theory of coverings of differential equations (see \cite{KV,KVi} and references therein) 
and several examples were given by Kiso \cite{Ki} about that time. However, it appears that it 
is only in \cite{Ga,llst,s-pre} that they have been used to find explicit solutions. It is then 
of interest to further study the work carried out in these three papers, and to show 
novel applications of the theory. This is what the geometric constructs of this article allow 
one to do:

The notion of a scalar equation describing pseudo-spherical 
surfaces (or ``of pseudo-spherical type'') is introduced in Section 2. Equations in this class
are of interest because they share with the sine--Gordon equation the property that their 
(suitably generic) solutions determine two--dimensional surfaces equipped with Riemannian 
metrics of constant Gaussian curvature $-1$, and also because equations possessing this structure
are naturally the integrability condition of $sl(2,{\bf R})$--valued linear problem. Section 3
is on (local/nonlocal) symmetries and pseudo-potentials for equations describing pseudo-spherical
surfaces. A short introduction to the theory of coverings is also included here: it appears in 
Subsection 3.2. From the point of view of this paper, the pseudo-potentials of the scalar
equations considered in \cite{Ga,llst,s-pre} determine geodesics of the pseudo-spherical 
structures described by the equations at hand, and the nonlocal symmetries $G$ for them are
obtained by studying infinitesimal deformations $u \mapsto u + \tau\,u$ of the dependent variable
$u$ which preserve geodesics to first order in the deformation parameter $\tau$.

Section 4 contains the application of the work carried out in Sections 2 and 3 to the the 
Korteweg--de Vries \cite{KdV}, Camassa--Holm \cite{CH} and Hunter--Saxton \cite{HS,HZ} equations.
Their zero curvature representations, quadratic pseudo-potentials, and modified versions are
introduced, and then nonlocal symmetries of ``pseudo-potential type'' are constructed for them. 
Furthermore, it is shown that these symmetries can be integrated, and that consideration of their
flows yields smooth local existence theorems for solutions. Examples of solutions are also 
included here. 

Special cases of some of the results appearing in this paper have been announced in \cite{R4,R5}.

\section{Equations of pseudo-spherical type}

Equations of pseudo-spherical type were introduced by S.S. Chern and K. Tenenblat in 1986 
\cite{ChT}, motivated by the fact that \cite{S} generic solutions of equations integrable by 
the Ablowitz, Kaup, Newell and Segur (AKNS) inverse scattering scheme determine --whenever their
associated linear problems are real-- pseudo-spherical surfaces, that is, Riemannian surfaces of
constant Gaussian curvature $-1$.

Below and henceforth, partial derivatives
$\displaystyle \frac{\partial^{p+q}u}{\partial x^{p}\partial t^{q}}$ are denoted by 
$u_{x^{p}t^{q}}$.

\begin{definition}         \label{epss}
A scalar differential equation $\Xi (x, t, u, u_{x}, \dots, u_{x^{n}t^{m}}) = 0$ in two 
independent variables $x,t$ is of pseudo-spherical type (or, it describes 
pseudo-spherical surfaces) if there exist one-forms $\omega^{i} \neq 0$, $i =  1,2,3$,
\begin{equation}   \label{omega}
\omega^{i} = f_{i 1}(x, t, u, \dots, u_{x^{r}t^{p}})\,dx + 
             f_{i 2}(x, t, u, \dots, u_{x^{s}t^{q}})\,dt \; , \; \; \; \; \; 
\end{equation}
whose coefficients $f_{ij}$ are differential functions, such that the one--forms
$\overline{\omega}^{i} = \omega^{i}(u(x,t))$  satisfy the structure equations 
\begin{equation}               \label{structure0}
d\,\overline{\omega}^{1} = \overline{\omega}^{3} \wedge \overline{\omega}^{2} \; ,
\; \; \; \; \; \; \; \; \; 
d\,\overline{\omega}^{2} = \overline{\omega}^{1} \wedge \overline{\omega}^{3} \; ,
\; \; \; \; \; \; \; \; \; 
d\,\overline{\omega}^{3} = \overline{\omega}^{1} \wedge \overline{\omega}^{2} \; ,
\end{equation}
whenever $u=u(x,t)$ is a solution to $\Xi = 0$.
\end{definition}

Recall that a {\em differential function} is a smooth function depending on the independent 
variables $x,t$, the dependent variable $u$, and a finite number of derivatives of $u$, see 
\cite{O}. The trivial case when all the functions $f_{ij}$ 
depend only on $x$ and $t$ is excluded from the considerations below.

\begin{example}
The equation
\begin{equation}
- f_{t} + \frac{\partial}{\partial x}\,[ g_{x} + f g ] = 0 \; , \label{geo}
\end{equation}
in which $f$ and $g$ are arbitrary differential functions, is of pseudo-spherical type
with associated one--forms
\[
\omega^{1} = f dx + ( g_{x} + f\,g ) dt \; , \; \; \; \; \; \; 
\omega^{2} = \lambda\,dx + \lambda g dt \; , \; \; \; \; \; \; 
\omega^{3} = - \lambda\,dx - \lambda g dt \; .
\]
The well--known Burgers equation $u_{t} = u_{xx} + u u_{x}$, is a special case of (\ref{geo})
with $f = g = (1/2) u$. 
\end{example}

The expression ``PSS equation'' is sometimes used in this paper instead of ``equation of 
pseudo-spherical type''. The geometric interpretation of Definition \ref{epss} is based on the 
following genericity notions (\cite{R1.5} and references therein):

\begin{definition}
Let $\Xi = 0$ be a PSS equation with associated one--forms
$\omega^{i}$, $i=1,2,3$. A solution $u(x,t)$ of $\Xi = 0$ is $I$--generic if
$(\omega^{3}\wedge\omega^{2})(u(x,t)) \neq 0$, $II$--generic if
$(\omega^{1}\wedge\omega^{3})(u(x,t)) \neq 0$, and $III$--generic if
$(\omega^{1}\wedge\omega^{2})(u(x,t)) \neq 0$.
\end{definition} 

\begin{proposition} \label{p1}
Let $\Xi = 0$ be a PSS equation with associated one--forms $\omega^{i}$.

{\em (a)} If $u(x,t)$ is a $I$--generic solution, $\overline{\omega}^{2}$ and 
$\overline{\omega}^{3}$ determine a Lorentzian metric of Gaussian curvature $K = -1$ on the 
domain of $u(x,t)$, with metric connection one--form $\overline{\omega}^{1}$.

{\em (b)} If $u(x,t)$ is a $II$--generic solution, $\overline{\omega}^{1}$ and 
$- \overline{\omega}^{3}$ determine a Lorentzian metric of Gaussian curvature $K = -1$ on 
the domain of $u(x,t)$, with metric connection one--form $\overline{\omega}^{2}$.

{\em (c)} If $u(x,t)$ is a $III$--generic solution, $\overline{\omega}^{1}$ and 
$\overline{\omega}^{2}$ determine a Riemannian metric of Gaussian curvature $K = -1$ on 
the domain of $u(x,t)$, with metric connection one--form $\overline{\omega}^{3}$. 
\end{proposition}

Proposition \ref{p1} follows from the structure equations of a (pseudo) Riemannian
manifold, which appear, for example, in \cite{T}. The notion of integrability 
introduced below is implicit in \cite{ChT}. 

\begin{definition}   \label{gi}
An equation is geometrically integrable if it describes a 
non--trivial one--parameter family of pseudo-spherical surfaces.        
\end{definition} 

\begin{proposition} 
A geometrically integrable equation $\Xi = 0$ with associated one--forms $\omega^{i}$, 
$i = 1,2,3$, is the integrability condition of a one--parameter family of $sl(2,{\bf R})
$--valued linear problems.
\end{proposition}
\begin{proof}
The linear problem $d \psi = \Omega \psi$, in which 
\begin{equation}    \label{linear0}
\Omega  = X dx + T dt    = \frac{1}{2} \left(
                         \begin{array}{lr}
                                 \omega ^{2}& \omega ^{1} - \omega ^{3}\\
                                 \omega ^{1} + \omega ^{3}&  - \omega ^{2}
                          \end{array} \right),
\end{equation} 
is integrable whenever $u(x,t)$ is a solution of $\Xi = 0$. 
\end{proof}

An important idea in integrable systems \cite{FT} is that an equation $\Xi = 0$ is not just the 
integrability condition of a linear problem $\psi_{x} = X \psi$, $\psi_{t} = T \psi$, but that 
the zero curvature equation $X_{t} - T_{x} + [X,T] = 0$ is {\em equivalent} to $\Xi =0$. It is 
a crucial problem to formalize this remark within the context of PSS equations. For evolutionary
equations $u_{t} = F(x,t,u, ... , u_{x^{n}})$ one proceeds thus \cite{KT, R1.5}: 

Consider the differential ideal $I_{F}$ generated by the two--forms
\[
du \wedge dx + F(x,t,u, ... , u_{x^{n}}) dx \wedge dt \; , \; \; \; \; 
du_{x^{l}} \wedge dt - u_{x^{l+1}}\, dx \wedge dt \; , \; \; \; \; \; \;  1 \leq l \leq n-1 \, ,
\]
on a manifold $J$ with coordinates $x,t,u,u_{x},\ldots ,u_{x^{n}}$. 

\begin{definition} \label{nk}
An evolution equation $ u_{t} = F(x,t,u, ... , u_{x^{n}})$ is strictly pseudo-spherical if
there exist one--forms $\omega^{i} = f_{i 1}\,dx + f_{i 2}\,dt$, $i=1,2,3$, 
whose coefficients $f_{ij}$ are smooth functions on $J$, such that the two--forms
\begin{equation}
\Omega_{1} = d \omega^{1} - \omega^{3} \wedge \omega^{2} \; , \; \; \; \; 
\Omega_{2} = d \omega^{2} - \omega^{1} \wedge \omega^{3} \; , \; \; \; \;  
\Omega_{3} = d \omega^{3} - \omega^{1} \wedge \omega^{2} \; ,                   \label{mari7}
\end{equation}
generate $I_{F}$.
\end{definition}

Local solutions of $u_{t} = F$ correspond to integral submanifolds of the exterior 
differential system $\{ I_{F} , dx \wedge dt\}$. Thus, is $u_{t} = F$ is strictly 
pseudo-spherical, it is necessary and sufficient for the structure equations $\Omega_{\alpha} 
= 0$ to hold. The following lemma \cite{R1, R1.5} is used in Section 3 below.

\begin{lemma}   \label{lemma0}
Necessary and sufficient conditions for an $n^{\rm th}$ order equation $u_{t} = F$ to be 
strictly pseudo-spherical are the conjunction of:

{\em (a)} The functions $f_{ij}$ satisfy $f_{i 1, u_{x^{a}}} = 0$, $a \geq 1$; 
$f_{i 2,u_{x^{n}}} = 0$, $i = 1,2,3$; and 
\begin{equation}
f_{11,u}^{2} + f_{21,u}^{2} + f_{31,u}^{2} \neq 0 \;  ,                          \label{sve00}
\end{equation}

{\em (b)} $F$ and $f_{ij}$ satisfy the identities
\begin{equation}
\; \; \; \; \; \;  -f_{i 1,u}F + \sum_{p=0}^{n-1}u_{x^{p+1}}f_{i 2,u_{x^{p}}} +
f_{j 1}f_{k 2} - f_{k 1}f_{j 2} + f_{i 2,x} - f_{i 1,t} = 0 \; ,                  \label{sve0}
\end{equation}
in which $(i,j,k) \in \{(1,2,3),(2,3,1),(3,2,1) \}$. 
\end{lemma}

\section{Symmetries and pseudo-potentials for PSS equations}

\subsection{Pseudo-potentials}

The following geometrical result appears in \cite{ChT, T}:

\begin{proposition} 
Given an orthogonal coframe $\{ \overline{\omega}^{1}, \overline{\omega}^{2} \}$ and 
corresponding metric connection one-form $\overline{\omega}^{3}$ on a Riemannian surface $M$ 
with metric $\overline{\omega}^{1}\otimes\overline{\omega}^{1} + 
\overline{\omega}^{2}\otimes\overline{\omega}^{2}$, there exists a new orthogonal coframe 
$\{ \overline{\theta}^{1}, \overline{\theta}^{2} \}$ and new metric connection one-form 
$\overline{\theta}^{3}$ on $M$ satisfying
\begin{equation}
d \overline{\theta}^{1} = 0 \; , \; \; \; \; \; 
d \overline{\theta}^{2} = \overline{\theta}^{2} \wedge \overline{\theta}^{1} \; , \; \; \; \; \; 
\mbox{ and } \; \; \; \; \; 
\overline{\theta}^{3} + \overline{\theta}^{2}  = 0 \; ,                               \label{cl}
\end{equation}
if and only if the surface $M$ is pseudo-spherical. 
\end{proposition}
\begin{proof}
Assume that the local orthonormal frames dual to the coframes 
$\{ \overline{\omega}^{1}, \overline{\omega}^{2} \}$ and
$\{ \overline{\theta}^{1}, \overline{\theta}^{2}\}$ possess the same orientation. The one--forms 
$\overline{\omega}^{\alpha}$ and $\overline{\theta}^{\alpha}$ are then connected by means of 
\begin{equation}
\overline{\theta}^{1} =  \overline{\omega}^{1} \cos \rho +
\overline{\omega}^{2} \sin \rho \; ,                              \; \; \; \; \; \;           
\overline{\theta}^{2} =  - \overline{\omega}^{1} \sin \rho +
\overline{\omega}^{2} \cos \rho \; ,                             \; \; \; \; \; \;  
\overline{\theta}^{3} = \overline{\omega}^{3} + d \rho \; .                          \label{co3}
\end{equation}
It follows that one--forms $\overline{\theta}^{1}, \overline{\theta}^{2}, \overline{\theta}^{3}$ 
satisfying (\ref{cl}) exist if and only if the Pfaffian system
\begin{equation}
\overline{\omega}^{3} + d \rho - \overline{\omega}^{1} \sin \rho + 
\overline{\omega}^{2} \cos \rho  = 0                                             \label{ps0}
\end{equation}
on the space of coordinates $(x,t,\rho)$ is completely integrable for $\rho (x,t)$, and it is 
easy to see that this happens if and only if $M$ is pseudo-spherical.
\end{proof}
 
Equations (\ref{cl}) and (\ref{ps0}) determine geodesic coordinates on $M$ \cite{ChT,T}. If an 
equation $\Xi = 0$ describes pseudo-spherical surfaces with associated one-forms $\omega^{i} = 
f_{i 1}dx + f_{i 2}dt$, Equations (\ref{cl}) and (\ref{ps0}) imply that the Pfaffian 
system
\begin{equation}
{\omega}^{3}(u(x,t)) + d \rho - {\omega}^{1}(u(x,t)) \sin \rho + 
{\omega}^{2}(u(x,t)) \cos \rho  = 0                                                    \label{s}
\end{equation} 
is completely integrable for $\rho (x,t)$ whenever $u(x,t)$ is a local solution of $\Xi = 0$. 

\begin{remark}
Equations (\ref{cl}) and (\ref{co3}) imply that for each solution $u(x,t)$ and 
corresponding solution $\rho (x,t)$ of (\ref{s}), the one-form 
\begin{equation}
\theta^{1}(u(x,t)) = {\omega}^{1}(u(x,t)) \cos \rho + {\omega}^{2}(u(x,t)) \sin \rho \label{clo}
\end{equation} 
is closed. Since one--forms which are closed on solutions of $\Xi = 0$ determine conservation 
laws \cite{KV,KVi,O} it follows that if the functions $f_{ij}$ (and therefore $\rho (x,t)$ and 
$\theta^{1}$) can be expanded as power series in a parameter $\lambda$, the PSS equation 
$\Xi = 0$ will possess, in principle, an infinite number of conservation laws, which may well be 
nonlocal. The reader is referred to \cite{ChT,jmp,R1,R4,R5,T} for further discussions. 
\end{remark}

\begin{lemma}  \label{the1}
Let $\Xi = 0$ be a PSS equation with associated one--forms $\omega^{i}$. Under the 
changes of variables $\Gamma = \tan (\rho /2)$ and $ \hat{\Gamma} = \cot (\rho /2)$, the Pfaffian
system $(\ref{s})$ and the one--form $(\ref{clo})$ become, respectively, 
\begin{eqnarray}
- 2 d \, \Gamma & = & (\overline{\omega}^{3} + \overline{\omega}^{2}) - 2 \Gamma
\overline{\omega}^{1} + \Gamma^{2}(\overline{\omega}^{3} - \overline{\omega}^{2})\; ,
                                                                                \label{sgamma} \\
\Theta & = & \overline{\omega}^{1} - \Gamma (\overline{\omega}^{3} - \overline{\omega}^{2}) \; ,
  \; \; \; \; \; \; \; \; \mbox{ $($up to an exact differential form$)$}  \label{alla1} \\
\mbox{ and } \; \; \; \; \; \; \; & & \nonumber \\
2 d \, \hat{\Gamma} & = & (\overline{\omega}^{3} - \overline{\omega}^{2}) - 2 \hat{\Gamma}
\overline{\omega}^{1} + \hat{\Gamma}^{2}(\overline{\omega}^{3}+\overline{\omega}^{2}) \; ,
                                                                               \label{sgamma2} \\
\hat{\Theta} & = & - \overline{\omega}^{1} + \hat{\Gamma} (\overline{\omega}^{3} + 
\overline{\omega}^{2}) \; ,  \; \; \; \; \mbox{ $($up to an exact differential form$)$},
                                                                                   \label{alla2}
\end{eqnarray}
in which $\overline{\omega}^{i} = \omega^{i}(u(x,t))$, $i=1,2,3$.
\end{lemma}

Pseudo--potentials are defined as a generalization of conservation laws:

\begin{definition}
A real--valued function $\Gamma$ is a pseudo--potential of a differential equation $\Xi(x,t,u,
\dots ,u_{x^{m}t^{n}}) = 0$ if there exist smooth functions $f,g$ depending on $\Gamma$, $x,t,u$,
and a finite number of derivatives of $u$, such that the one--form 
\[
\Omega_{\Gamma} = d \Gamma - ( f dx + g dt )
\]
satisfies
\[
d \Omega_{\Gamma} = 0 \; \; \; \; \mbox{ mod } \Omega_{\Gamma}
\]
whenever $u(x,t)$ is a solution to $\Xi = 0$. 
\end{definition}

One says that the one--form $\Omega_{\Gamma}$ is {\em associated} to the pseudo--potential 
$\Gamma$. Note that if the functions $f,g$ appearing in the definition do not depend on $\Gamma$,
$\Omega_{\Gamma}$ is a {\em bona fide} conservation law of the equation $\Xi = 0$.

Pseudo--potentials were introduced by Wahlquist and Eastbrook \cite{WE}. They can be understood 
geometrically in the framework of covering theory, see \cite{KV,KVi} and references therein. 
{\em Quadratic} pseudo--potentials, that is, pseudo-potentials $\Gamma$ such that the functions 
$f,g$ appearing in the associated one--form $\Omega_{\Gamma}$ are quadratic polynomials in 
$\Gamma$, possess a very appealing geometrical interpretation within the framework of PSS 
equations.

\begin{proposition}
A differential equation $\Xi = 0$ is of pseudo-spherical type if and only if it admits a
quadratic pseudo--potential.
\end{proposition}
\begin{proof}
Equations (\ref{sgamma}) and (\ref{sgamma2}) say that $\Gamma$ and $\hat{\Gamma}$ are 
pseudo--potentials of Riccati type for the PSS equation $\Xi = 0$. On the other hand, if
$\Xi = 0$ admits a pseudo--potential $\Gamma$ with associated one--form
$\Omega_{\Gamma} = d \Gamma - ( f dx + g dt )$, in which $f=a+b\Gamma + c\Gamma^{2}$ and
$g=a'+b'\Gamma + c'\Gamma^{2}$, the system
\begin{equation}
\Gamma_{x} = a + b \Gamma + c \Gamma^{2} \; , \; \; \; \; \;
\Gamma_{t} = a' + b' \Gamma + c' \Gamma^{2} \; ,              \label{qpp}
\end{equation}
is completely integrable on solutions of $\Xi = 0$, and therefore the equations
\[
a_{t} + b a' = a'_{x} + b'a \; , \; \; \; \; \; b_{t} + 2 c a' = b'_{x} + 2 c'a 
\; , \; \; \; \; \; c_{t} + c b' = c'_{x} + c'b 
\]
are satisfied on solutions of $\Xi = 0$. It follows that $\Xi = 0$ is a PSS equation with
associated one--forms $\omega^{i}$, $i=1,2,3$, given by 
\begin{equation}
\omega^{1} =  b dx + b' dt \; , \; \; \; \; 
\omega^{2} = (-a + c ) dx + (-a' + c') dt \; , \; \; \; \; 
\omega^{3} = -(a+c) dx - (a' + c' ) dt \; .                \label{of}
\end{equation} 
\end{proof}

The quadratic pseudo-potential (\ref{sgamma}) induced by the one--forms (\ref{of}) is,
of course, (\ref{qpp}), and therefore if a differential equation $\Xi = 0$ admits a quadratic 
pseudo--potential $\Gamma$, the function $\Gamma$ determines the geodesics of the 
pseudo--spherical structures described by $\Xi = 0$.

\subsection{Symmetries}

As stated in Section 1, a differential function $G$ is a generalized symmetry of $u_{t}=F$ if 
and only if $u(x,t) + \tau G(u(x,t))$ is --to first order in $\tau$-- a solution of $u_{t}=F$ 
whenever $u(x,t)$ is a solution of $u_{t}=F$. In other words, $G$ is a generalized symmetry of
$u_{t} = F$ if and only if the equation $D_{t} G = F_{\ast} G$, in which $F_{\ast}$ denotes the 
formal linearization of $F$,
\begin{equation}
F_{\ast} = \sum_{i=0}^{k} \frac{\partial F}{\partial u_{x^{i}}} \, D_{x}^{i} \; ,
\end{equation}
and $D_{x}$, $D_{t}$ are the total derivative operators with respect to $x$ and $t$ respectively
\cite{MSS,O}, holds identically once all the derivatives with respect to $t$ appearing in it have
been replaced by means of $u_{t}=F$. This definition extends straightforwardly to (systems of)
equations not necessarily of evolutionary type, see \cite{O,KVi}.

Now, let $u_{t} = F$ be an $n^{\rm th}$ order strictly pseudo-spherical evolution equation with 
associated one--forms $\omega^{i}$. Let $u(x,t)$ be a local solution of $u_{t} = F$, and set 
$\overline{G} = G(u(x,t))$, in which $G$ is a differential function. Expand 
$\omega^{i}(\,u(x,t) + \overline{G}\,)$ about $\tau=0$, thereby obtaining an infinitesimal
deformation 
$\overline{\omega}^{i} + \tau \overline{\Lambda}_{i}$, $\overline{ \Lambda}_{i} = 
\overline{g}_{i 1}dx + \overline{g}_{i 2}dt$, of the one--forms 
$\overline{\omega}^{i} = \omega^{i}(u(x,t))$. Lemma 1 implies that 
$\overline{g}_{i 1} = 
f_{i 1 ,u}(u(x,t)) \,\overline{G}$ and $\overline{g}_{i 2} = \sum_{p=0}^{n-1} 
f_{i 2,u_{x^{p}}}(u(x,t))(\partial^{p}\overline{G}/\partial x^{p})$, $i=1,2,3$.
One then has \cite{R1}:

\begin{theorem}   \label{prop1}
Suppose that $u_{t} = F(x,t,u, \dots u_{x^{n}})$ is strictly pseudo-spherical with associated 
one-forms $\omega^{i} = f_{i 1}dx + f_{i 2}dt$, $i = 1,2,3$, and let $G$ be 
a differential function. The deformed one--forms $\overline{\omega}^{i} + \tau 
\overline{\Lambda}_{i}$ satisfy the structure equations of a pseudo-spherical surface up to
terms of order $\tau^{2}$ if and only if $G$ is a generalized symmetry of $u_{t} = F$.
\end{theorem}

Thus, generalized symmetries of strictly pseudo-spherical equations $u_{t}=F$ are identified with
infinitesimal deformations of the pseudo-spherical structures determined by $u_{t}=F$ which 
preserve Gaussian curvature to first order in the deformation parameter. Theorem 
\ref{prop1} has been used in \cite{R1, R1.5} to show the existence of (generalized, 
nonlocal) symmetries of strictly PSS equations of evolutionary type. 

The symmetry concept is now extended to encompass nonlocal data \cite{KV,KVi}. In order to do 
this, one needs some notions from the geometric theory of differential equations
\cite{KV,KVi,O,jmp}: 

Let $E$ be a trivial bundle given locally by $(x,t,u) \mapsto (x,t)$, and let $J^{\infty}E$ be 
the corresponding infinite jet bundle of $E$. Then, 

(a) A scalar differential equation $\Xi = 0$ in two independent variables $x,t$ is 
identified with a sub--bundle $S^{\infty}$ of $J^{\infty}E$ called the {equation manifold} of 
$\Xi = 0$. 

(b) The fiber bundles $S^{\infty}$ and $J^{\infty}E$ come equipped with flat connections 
---the {\em Cartan connections} of $S^{\infty}$ and $J^{\infty}E$--- which agree on $S^{\infty}$.
 
(c) The horizontal vector fields of these 
connections are (locally) linear combinations of the total derivatives $D_{x}$ and $D_{t}$.

\begin{definition}  \label{cov}
Let $\Xi = 0$ be a differential equation with equation manifold $S^{\infty}$, and let 
$\pi : \overline{S} \rightarrow S^{\infty}$ be a fiber bundle over $S^{\infty}$. 
The bundle $\pi$ determines a covering structure (or, $\overline{S}$ is a covering of 
$S^{\infty}$) if and only if 

(a) There exists a flat connection $\overline{C}$ on the bundle 
$\pi^{\infty}_{M} \circ \pi : \overline{S} \rightarrow M$, and 

(b) The connection $\overline{C}$ agrees with the Cartan connection $C$ on $S^{\infty}$, that is,
for any vector field $X$ on $M$, $\pi_{\ast} (\overline{X}) = pr^{\infty}(X)$, in which the 
vector field $\overline{X}$ on $\overline{S}$ is the horizontal lift of $X$  induced by 
$\overline{C}$, and $pr^{\infty}(X)$ is the horizontal lift of the vector field $X$ with respect
to the Cartan connection of $S^{\infty}$.
\end{definition}

Fiber bundles over equations manifolds $S^{\infty}$ are rigorously defined in \cite{KV,KVi}.
Consider local coordinates $(x, t, u, \dots ,w^{1}, \dots , w^{N} )$, 
$1 \leq N \leq \infty$, on a covering $\pi : \overline{S} \rightarrow S^{\infty}$ of 
$S^{\infty}$ such that $(x,t,u,\dots)$ are canonical coordinates on $S^{\infty}$ and 
$(w^{1},\dots ,w^{N})$ are fiber coordinates on $\overline{S}$, and let $D_{x}$ and $D_{t}$ 
be the total derivative operators on $S^{\infty}$. Definition \ref{cov} implies that 
the covering $\overline{S}$ is determined locally by the data $(\overline{S}, \overline{D}_{x}, 
\overline{D}_{t} , \pi )$, in which $\overline{D}_{x}$, $\overline{D}_{t}$ are differential 
operators on $\overline{S}$ satisfying:

(a) $\overline{D}_{x}$, $\overline{D}_{t}$ are of the form
\begin{equation}
\overline{D}_{x}=D_{x} + X_{1} \; \; \; \; \mbox{ and } \; \; \; \; 
\overline{D}_{t}=D_{t}+X_{2} \; ,                                                   \label{cdete}
\end{equation}
in which $X_{i}$, $i = 1, 2$, are vertical vector fields on $\overline{S}$,  
$X_{i} = \sum_{\beta = 1}^{N} X_{~i}^{\beta}\,\partial/\partial w^{\beta}$, and 

(b) $\overline{D}_{x}$ and $\overline{D}_{t}$ satisfy the integrability condition
\begin{equation}
[\,\overline{D}_{x} , \overline{D}_{t}\,] := D_{x}(X_{2}) - D_{t}(X_{1}) 
+ [ X_{1} , X_{2} ] = 0 \; .                                                      \label{cove}
\end{equation}

The operators $\overline{D}_{x}$ and $\overline{D}_{t}$  are the {\em total derivative operators}
on $\overline{S}$. As in the case of total derivatives on equation manifolds, $\overline{D}_{x}$
and $\overline{D}_{t}$ span the horizontal distribution of $\overline{S}$. The fiber coordinates
$w^{i}$, $1 \leq i \leq N$, are called {\em nonlocal variables} with respect to $S^{\infty}$, 
and $N$ is the {\em dimension} of the covering $\pi : \overline{S} \rightarrow S^{\infty}$. 

\begin{example}
Assume that the one--form $\kappa = f dx + g dt$, in which $f$ and $g$ are
differential functions, satisfies $D_{t}f = D_{x}g$ on solutions of $u_{t} = F$. Then, 
$\kappa$ determines a one--dimensional covering $(\overline{S}, \overline{D}_{x}, 
\overline{D}_{t}, \pi)$ of $u_{t} = F$: $\overline{S}$ is locally defined by $\overline{S}  
= \{ (x,t,u, \dots ,u_{x^{m}}, \dots , w) \}$, where $(x,t,u, u_{x}, ...)$ are coordinates on
the equation manifold $S^{\infty}$ of $u_{t} = F$, and
\begin{equation}
\overline{D}_{x} = D_{x} + f \frac{\partial}{\partial w}\; , \; \; \; \; \; \; 
\overline{D}_{t} = D_{t} + g \frac{\partial}{\partial w}\; .
\end{equation} 
It is trivial to check that $D_{t}f = D_{x}g$ implies that the integrability condition 
(\ref{cove}) for $\overline{D}_{x}$ and $\overline{D}_{t}$ holds.
\end{example}

\begin{example}
Let $S^{\infty}$ be the equation manifold of the ``trivial'' equation, that is,
$S^{\infty} = J^{\infty}E$. Set $u_{k_{1},k_{2}} = D^{k_{1}}_{x}D^{k_{2}}_{t}u$, where 
$k_{1},k_{2} \in {\bf Z}$, and $u_{0,0} = u$. Introduce the manifold $\overline{S}$ locally by
$\overline{S} = \{ (x,t,,u, \dots, u_{k_{1},k_{2}}, \dots ) \}$ and define the projection map 
$\pi : \overline{S} \rightarrow S^{\infty}$ in an obvious way. For any pair 
$(k_{1},k_{2}) \in {\bf Z}^{2}$, let $\pi_{k_{1},k_{2}}$ be the function
\[
\pi_{k_{1},k_{2}} = \left\{
                 \begin{array}{ccc}
 \displaystyle   \frac{x^{-k_{1}}t^{-k_{2}}}{(-k_{1}\!)(-k_{2}\!)} & & k_{1},k_{2} \leq 0 \; ,\\
                 0                                                 & & \mbox{otherwise} \; .
                 \end{array}
                    \right. 
\]
The {\em ghost vector fields} $\gamma_{k_{1},k_{2}}$, where $(k_{1},k_{2}) \in 
{\bf Z}^{2}$, are defined by the rules
\[
\gamma_{k_{1},k_{2}}(u_{m,n}) = \pi_{k_{1} + m , k_{2} + n} \; , \; \; \; \; \; \;
\gamma_{k_{1},k_{2}}(x^{i}t^{j}) = 0 \; .
\]
Ghost vector fields have been quite recently introduced by Olver, Sanders, and Wang \cite{OSW},
as a way to extend the Lie bracket of evolutionary vector fields to the nonlocal domain.
Now set $X_{1} = \sum \gamma_{k,0}$, and $X_{2} = \sum \gamma_{0,k}$. Then, $(\overline{S}, 
\overline{D}_{x}, \overline{D}_{t}, \pi)$ in which $\overline{D}_{x} = D_{x} + X_{1}$ and
$\overline{D}_{t} = D_{t} + X_{2}$, is an infinite dimensional covering of $S^{\infty}$.
That the integrability condition (\ref{cove}) holds, follows from the fact that ghost vector
fields commute with each other, see \cite{OSW}. 
\end{example}

Nonlocal symmetries are defined thus:

\begin{definition}
Let $(\overline{S}, \overline{D}_{x}, \overline{D}_{t}, \pi )$ be an $N$--dimensional covering 
of the $n^{\rm th}$ order equation $u_{t} = F$ equipped with coordinates 
$(x, t, u, u_{x}, \dots , w^{\beta})$, $1 \leq \beta \leq N$, and assume that $\overline{D}_{x}$
and $\overline{D}_{t}$ are given by 
\begin{equation}
\overline{D}_{x}=D_{x} + \sum_{\beta = 1}^{N} X_{~1}^{\beta}\,\frac{\partial}{\partial w^{\beta}}
\; \; \; \; \mbox{ and } \; \; \; \; 
\overline{D}_{t}=D_{t} + \sum_{\beta = 1}^{N} X_{~2}^{\beta}\,\frac{\partial}{\partial w^{\beta}}
                  \; .                                                              \label{cdxdt}
\end{equation}
A nonlocal symmetry of type $\pi$ of $u_{t} = F$ is a vector field $\overline{D}_{\tau}$ of the 
form
\begin{equation}
\overline{D}_{\tau} = \sum_{i=0}^{\infty}\overline{D}_{x}^{i}(G)\frac{\partial}{\partial
u_{x^{i}}} + \sum_{\beta} I_{\beta} \frac{\partial}{\partial w^{\beta}} \; ,
\end{equation}
in which $G$ and $I_{\beta}$, $1 \leq \beta \leq N$, are smooth functions on
$\overline{S}$, such that the following equations hold:
\begin{equation}
\overline{D}_{t} G = \sum_{i=0}^{n}\frac{\partial F}{\partial u_{x^{i}}}\overline{D}_{x}^{~i}(G),
\; \; \; \; \; \; \; 
\overline{D}_{x} (I_{\beta}) = \overline{D}_{\tau} (X^{\beta}_{1}), \; \; \; \; \; \; \; 
\overline{D}_{t} (I_{\beta}) = \overline{D}_{\tau} (X^{\beta}_{2}) \; .           \label{shadow}
\end{equation}
\end{definition}

More generally, if $\Xi = 0$  is a scalar differential equation in two independent variables
$x,t$, not necessarily of evolutionary type, a nonlocal symmetry of type $\pi$ of $\Xi = 0$ is 
a vector field
\begin{equation}
\overline{D}_{\tau} = \sum \overline{D}_{x}^{~i}\overline{D}_{t}^{~j}(G)
                      \frac{\partial}{\partial u_{x^{i}t^{j}}}
                      + \sum_{\beta} I_{\beta} \frac{\partial}{\partial w^{\beta}} \;  ,
\end{equation}
where $u_{x^{i}t^{j}}$ denote intrinsic coordinates on the equation
manifold of $\Xi = 0$, such that 
\begin{equation}
\overline{\Xi}_{\ast}(G) = 0 \; , \; \; \; \; \; \; \; 
\overline{D}_{x} (I_{\beta}) = \overline{D}_{\tau} (X^{\beta}_{1}) \; , \; \; \; \; \; \; \; 
\overline{D}_{t} (I_{\beta}) = \overline{D}_{\tau} (X^{\beta}_{2}) \; ,           \label{shadow2}
\end{equation}
in which $\overline{\Xi}_{\ast}$ is the lift of the formal linearization of $\Xi$ to the
covering $\pi$,
\begin{equation}
\overline{\Xi}_{\ast} = \sum \frac{\partial\,\Xi}{\partial u_{x^{i}t^{j}}}\,\overline{D}_{x}^{~i}
 \, \overline{D}_{t}^{~j} \; .   \label{clinear}
\end{equation}

This definition can be adapted straightforwardly to systems of equations \cite{KV,KVi}, and this 
extension will be used in what follows without further ado.

Note that the first equations of (\ref{shadow}) and (\ref{shadow2}) depend only on $G$ and the
equation at hand. The vector field $G\,\partial/\partial u$, or, in the case of an
evolution equation
\begin{equation}
\sum_{i=0}^{\infty}\overline{D}_{x}^{i}(G)\frac{\partial}{\partial u_{x^{i}}} \; ,
                                                                                  \label{lieder}
\end{equation}
can be interpreted as a vector field on $\overline{S}$ {\em along} $S^{\infty}$. This vector 
field, or simply $G$, is called the {\em shadow} of the nonlocal symmetry $\overline{D}_{\tau}$.
In general, vector fields on $\overline{S}$ along $S^{\infty}$ which satisfy the first equation 
of (\ref{shadow}) are called {\em $\pi$--shadows}. An important question is whether one can 
extend $\pi$--shadows to {\em bona fide} nonlocal symmetries. General theorems along these lines 
have been proven by Nina Khor'kova in 1988, see \cite{KV,KVi}, and by Kiso \cite{Ki}. An example
of such an extension appears in Subsection 4.3 below.

Now one would like to characterize nonlocal symmetries of strictly pseudo-spherical evolution 
equations. Let $u_{t} = F$ be an $n^{\rm th}$ order strictly pseudo-spherical equation with 
associated one--forms $\omega^{i}$, $i = 1,2,3$, and equation manifold $S^{\infty}$, and consider
a covering $(\overline{S},\overline{D}_{x},\overline{D}_{t},\pi)$ of $S^{\infty}$. One first 
extends the ``horizontal'' exterior derivative operator from $S^{\infty}$ to $\overline{S}$
thus \cite{KVi}:

If $\omega = \sum a_{i_{1} \dots i_{k}}\,dx_{i_{1}} \wedge \dots \wedge dx_{i_{k}}$ is an 
horizontal differential form on $\overline{S}$, in which $x_{1} = x$, $x_{2} = t$, then
\[
\overline{d}_{H} \omega = \sum \left( \overline{D}_{x}a_{i_{1} \dots i_{k}}dx + 
         \overline{D}_{t}a_{i_{1} \dots i_{k}}dt \right) \wedge dx_{i_{1}} 
                                                              \wedge \dots \wedge dx_{i_{k}}\;.
\]
Next, let $G$ be a function on $\overline{S}$. In analogy with the generalized symmetry case, 
one studies the one--forms ${\omega}^{i} + \tau {\Lambda}_{i}$ on $\overline{S}$, in 
which ${ \Lambda}_{i} = {g}_{i 1}dx + {g}_{i 2}dt$ and
\begin{equation}
{g}_{i 1} = f_{i 1 ,u} \,{G}, \; \; \; \; \; \mbox{ and } \; \; \; \; \; 
{g}_{i 2} = \sum_{k=0}^{n-1} f_{i 2,u_{x^{k}}} \overline{D}_{x}^{~k} G, 
\; \; \; \; \; \; \; \;                                          i =  1,2,3.  \label{ling1}
\end{equation}

\begin{theorem}   \label{prop2}
Let $u_{t} = F(x,t,u, \dots u_{x^{n}})$ be strictly pseudo-spherical with associated one-forms
$\omega^{i} = f_{i 1}dx + f_{i 2}dt$, $i = 1,2,3$. Let $G$ be a smooth function on a covering 
$(\overline{S},\overline{D}_{x},\overline{D}_{t},\pi)$ of the equation manifold $S^{\infty}$, 
and consider the deformed one--forms ${\omega}^{\alpha} + \tau {\Lambda}_{\alpha}$ defined 
above. They satisfy the structure equations
\begin{equation}
\overline{d}_{H} \sigma^{1} = \sigma^{3} \wedge \sigma^{2}\; , \; \; \; \; \;
\overline{d}_{H} \sigma^{2} = \sigma^{1} \wedge \sigma^{3}\; , \; \; \; \; \;
\overline{d}_{H} \sigma^{3} = \sigma^{1} \wedge \sigma^{2}\; , \; \; \; \; \;  \label{steqn}
\end{equation}
up to terms of order $\tau^{2}$ if and only if $G$ is a $\pi$--shadow of the equation 
$u_{t} = F$.
\end{theorem}

\begin{proof}
The one--forms ${\omega}^{\alpha} + \tau {\Lambda}_{\alpha}$ satisfy (\ref{steqn}) up to terms 
of order $\tau^{2}$ if and only if
\begin{eqnarray}
-\overline{D}_{t}{g}_{11} + \overline{D}_{x}g_{12} & = & {f}_{31}
{g}_{22} - {f}_{32}{g}_{21} + {f}_{22}{g}_{31} - {f}_{21}{g}_{32},                 \label{g1} \\
-\overline{D}_{t}g_{21} + \overline{D}_{x}g_{22} & = &
{f}_{11}{g}_{32} - {f}_{12}{g}_{31} + {f}_{32}{g}_{11} - {f}_{31}{g}_{12}, \; \; \; \; \;
                                                                        \mbox{ and} \label{g2} \\
-\overline{D}_{t}g_{31} + \overline{D}_{x}g_{32} & = &
{f}_{11}{g}_{22} - {f}_{12}{g}_{21} + {f}_{22}{g}_{11} - {f}_{21}{g}_{12}.   \label{g3}
\end{eqnarray}

Since $u_{t} = F$ is strictly pseudo-spherical, Equations (\ref{sve0}) of Lemma \ref{lemma0} 
are identities. Take Lie derivatives with respect to the vector field $L_{\tau}$ defined in 
(\ref{lieder}), and substitute into (\ref{g1})--(\ref{g3}). One finds that these equations are 
satisfied if and only if
\begin{equation}
- {f}_{\alpha 1,u} \overline{D}_{t} (G) + {f}_{\alpha 1,u} 
\sum_{i=0}^{n} \frac{\partial \overline{F}}{\partial u_{x^{i}}}\, \overline{D}_{x}^{i}
({G}) =  0 , \; \; \; \; \; \; \; \; i = 1,2,3.
\end{equation}
Since the constraint (\ref{sve00}) holds, one concludes that Equations (\ref{g1})--(\ref{g3}) 
are satisfied if and only if $G$ is a$\pi$--shadow of the equation $u_{t} = F$.
\end{proof}

Theorem \ref{prop2} appeared for the first time in \cite{R1.5}; it is included here for 
ease of reference.

\section{Shallow water equations}

In this section the equations due to Korteweg and de Vries \cite{KdV},
\begin{equation}
u_{t} = u_{xxx} + 6 u u_{x}               \; ,                        \label{kdv}
\end{equation} 
Camassa and Holm \cite{CH},
\begin{equation}
m = u_{xx} - u \; , \; \; \; \; \; \; \; \; \; \; \; \; \; \; \; 
m_{t} = - m_{x}\,u - 2 \,m\,u_{x} \; ,                                  \label{ch}
\end{equation}
and Hunter and Saxton \cite{HS},
\begin{equation}
m = u_{xx} \; , \; \; \; \; \; \; \; \; \; \; \; \; \; \; \; 
m_{t} = - m_{x}\,u - 2 \,m\,u_{x} \; ,                                \label{hs}
\end{equation}
are studied taking advantage of the fact that they are members of a 
two--parameters family of equations of pseudo-spherical type. 

Of course, the KdV equation has been subject of an impressive body of research since \cite{KdV}, 
and in fact, Peter Olver has pointed out that (\ref{kdv}) was derived already in the 1870's by 
Boussinesq, who also found its first three conservation laws, and its one-soliton
and periodic traveling wave solutions, see \cite{Bou0,Bou}.

With respect to the integrability properties of the important Camassa--Holm \cite{CH} and 
Hunter--Saxton equations \cite{HS} the following (at least) is already 
known: their analysis by scattering/inverse scattering have been carried out (Beals, Sattinger 
and Szmigielski \cite{bss, bss1}), their bi-hamiltonian character has been discussed (Camassa 
and Holm \cite{CH}, Hunter and Zheng \cite{HZ}) and moreover, it has been proven that the 
Korteweg--de Vries, Camassa--Holm and Hunter--Saxton equations exhaust, in a precise sense, 
the bi-hamiltonian equations which can be modeled as geodesic flows on homogeneous spaces 
related to the Virasoro group (Khesin and Misio{\l}ek \cite{KM}). 

\subsection{Pseudo-spherical structures}

That the KdV equation describes pseudo-spherical surfaces was observed by Sasaki 
\cite{S} and also by Chern and Tenenblat, who obtained this result from
some general classification theorems proven by them in \cite{ChT}.

\begin{example}  \label{ej}
The KdV equation $u_{t} = u_{xxx} + 6 u u_{x}$ describes pseudo-spherical surfaces \cite{S, ChT}
with associated one--forms $\omega^{i} = f_{i 1}dx + f_{i 2}dt$, in which
\begin{eqnarray}
\omega^{1} & = & (1 - u)\,dx + (- u_{xx} + \lambda u_{x} - \lambda^{2} u - 2 u^{2} + 
                                                        \lambda^{2} + 2 u)\,dt      \label{a60} \\
\omega^{2} & = & \lambda\,dx + (\lambda^{3} + 2 \lambda u - 2 u_{x})\,dt \\
\omega^{3} & = & (- 1 - u)\,dx +(-u_{xx} +\lambda u_{x} - \lambda^{2} u - 2 u^{2} - \lambda^{2} 
                  - 2 u)\,dt,                                                         \label{a6}
\end{eqnarray}                              
and $\lambda$ is an arbitrary parameter. 
\end{example}

The analysis carried out in Section 3 allows one to obtain the standard quadratic 
pseudo-potential for KdV:

\begin{example}
Consider the KdV equation $u_{t} = u_{xxx} + 6 u u_{x}$, and the associated one--forms
$\omega^{i}$ given by (\ref{a60})--(\ref{a6}). Rotate the coframe $\{ \omega^{1},\omega^{2} \}$ 
determined by (\ref{a60})--(\ref{a6}) in $\pi/2$, and change $\Gamma$ for $- \Gamma$. One can 
then write the Pfaffian system (\ref{sgamma}) as 
\[
\mbox{(a)} \; \; \Gamma_{x} = - u - \lambda\Gamma - \Gamma^{2} \; , \; \; \; \; \; \; \; \; \; \;
\mbox{(b)} \; \; \Gamma_{t} = (\Gamma_{xx} - 3 \Gamma^{2} \lambda - 2 \Gamma^{3})_{x} \; . 
\]
Since the KdV equation is strictly pseudo-spherical, $u(x,t)$ as determined by (a) solves KdV if 
$\Gamma (x,t)$ solves (b). One has thus recovered the Miura transformation and the modified KdV 
equation from geometrical considerations.
\end{example}                                                    

Henceforth $\epsilon$ will denote a real parameter which equals 1 for CH and 0 for HS. 
Theorem \ref{ch-hs} below first appeared in \cite{R5}. It is reproduced here since
its proof will be used momentarily.

\begin{theorem} \label{ch-hs}
The Camassa-Holm and Hunter--Saxton equations, $(\ref{ch})$ and $(\ref{hs})$ respectively, 
describe pseudo-spherical surfaces.
\end{theorem}
\begin{proof}
Consider one--forms $\omega^{i}$, $i = 1,2,3$, given by
\begin{eqnarray}
\omega^{1} & = & \left( m -\beta + \epsilon \,{\alpha}^{-2} (\beta - 1) \right) \, dx\nonumber\\
   & & \; \; \; \; \; \; \; \; \; \; \; \; \; \; \; \; \; \; \; 
+ \left (- u_{{x}}\beta\,\alpha^{-1} - \beta \, {\alpha}^{-2} - u\,m - 1 + u \beta + 
u_{x} \, {\alpha}^{-1} + {\alpha}^{-2} \right ) \, dt  \; ,                     \label{s1}    \\
\omega^{2} & = & \alpha\,dx + \left(-{\beta}\,{\alpha}^{-1}-\alpha\,u+{\alpha}^{-1}+u_{{x}}\right)
 \, dt \; ,                                                                          \label{s2} \\
\omega^{3} & = & \left (m + 1 \right ) \, d x  +
\left(\epsilon \, u\,\alpha^{-2}(\beta - 1) - u\,m + {\alpha}^{-2} + \frac{u_{x}}{\alpha} - u 
- {\frac {\beta}{{\alpha}^{2}}}-{\frac {u_{{x}}\beta}{\alpha}}\right) \, dt\; , \; \; \; \;
                                                                                     \label{s3}
\end{eqnarray}
in which $m = u_{xx} - u$ and the parameters $\alpha$ and $\beta$ are constrained by the relation
\begin{equation}
{\alpha}^{2}+{\beta}^{2}-1=\epsilon \left[\frac{\beta - 1}{\alpha}\right]^{2} .\label{constraint}
\end{equation}
It is not hard to check that the structure equations (\ref{structure0}) are satisfied whenever
the equation 
\begin{equation}
-2\,u_{{x}}u_{{xx}}+3\,u_{{x}}\epsilon\,u- u u_{xxx}+\epsilon\,u_{t}-u_{{xxt}} = 0 \;  
										 \label{general}
\end{equation} 
holds, and Equation (\ref{general}) becomes the Camassa--Holm equation (\ref{ch}) if 
$m = u_{xx} - u$ and $\epsilon =1$, and the Hunter--Saxton equation (\ref{hs}) if $m = u_{xx}$
and $\epsilon = 0$.
\end{proof}

In order to include the KdV equation into the picture, one applies the Galilei transformation
${\cal T} : (X,T,U) \mapsto (x,t,u)$ given by 
\begin{eqnarray}
x & = & {\frac {X}{{\nu}}}+{\frac {T}{{\nu}^{3}}}-{\frac {T}{\sqrt{\nu}}} \; , \\
t & = & {\frac {T}{\sqrt{\nu}}} \; , \\
u & = & {\frac {U}{\sqrt{\nu}}}+ \frac{1}{3}\,\frac{1}{{\nu}^{5/2}} - \frac{1}{3} \; ,
\end{eqnarray}
to Equation (\ref{general}) and the one--forms (\ref{s1})--(\ref{s3}):

\begin{cor}
The nonlinear equation
\begin{equation}
-2\,{\nu}^{2}U_{X}U_{XX}+3\,\epsilon\,U_{X}\,U - {\nu}^{2}U_{XXX}\,U + 
\frac{2}{3}\,U_{XXX}-\frac{2}{3}\,{\nu}^{5/2}U_{XXX}+\epsilon\,U_{T}-{\nu}^{2}U_{XXT} = 0 
										\label{gen-eqn}
\end{equation}
describes pseudo-spherical surfaces with associated one--forms ${\cal T}^{\ast}\omega^{i}$,
in which $\omega^{i}$, $i = 1,2,3$ are given by $(\ref{s1})$--$(\ref{s3})$. 
\end{cor}

Equation (\ref{gen-eqn}) does contain the KdV, CH, and HS equations as special cases, but it is
interesting to remark that the one--forms ${\cal T}^{\ast}\omega^{i}$ are singular in the KdV 
limit $\nu \rightarrow 0$. For example, ${\cal T}^{\ast}\omega^{2}$ is 
\[
{\cal T}^{\ast}\omega^{2} =  
   \frac{\alpha}{\nu} \, d\,X
 + \left (-{\frac {\beta}{\sqrt{\nu}\,\alpha}}
-{\frac {\alpha\,U}{{\nu}}}+\frac{2}{3}\,{\frac {\alpha}{{\nu}^{3}}}-
\frac{2}{3}\,{\frac {\alpha}{\sqrt{\nu}}}+{\frac {1}{\sqrt{\nu}\,\alpha}}+U_{X}\right )\,d\,T\; .
\]
This difficulty is dealt with in the following subsection.

\begin{remark}
Equation (\ref{gen-eqn}) with $\epsilon =1$ and $1 - \nu^{5/2} = \gamma$ has been recently 
derived as a shallow water equation by Dullin, Gottwald, and Holm \cite{DGH}, via an asymptotic 
expansion of the Euler equations.
\end{remark}

\begin{remark}
Equation (\ref{gen-eqn}) can be interpreted as a geodesic equation on the Virasoro group.
In fact, (\ref{gen-eqn}) is in the class of equations studied by Khesin and Misio{\l}ek in
\cite{KM}: it is their equation (3.9) with $\beta = \nu^{2}$, $\alpha = \epsilon$, and
$b = (2/3)(1 - \nu^{5/2})$. 
\end{remark}

To continue the investigation, one dispenses with the constraint (\ref{constraint}) by using a 
parameterization of the curve ${\alpha}^{2}+{\beta}^{2}-1=\epsilon\,[(\beta - 1)/\alpha]^{2}$. 
For example one can take
\begin{equation}
\alpha = \sqrt{\epsilon + 1 - s^{2}} \; , \; \; \; \; \; \; \; \; \; \; \; \; \;
\beta = \frac{\epsilon}{s-1} - s \; .  \label{p-beta}
\end{equation}

After rotating by $\pi/2$ and using (\ref{p-beta}), the one--forms 
${\cal T}^{\ast}\omega^{i}$ associated with Equation (\ref{gen-eqn}) become:
\begin{eqnarray}
{\cal T}^{\ast}\omega^{1} & = &
{\frac {\sqrt {\epsilon+1-{s}^{2}}}{\nu}}dX
+ \left(\frac{2}{3}\,\sqrt{\epsilon +1-s^{2}}\,(\,\frac{1}{\nu^{3}}-\frac{s+1/2}{\sqrt{\nu}(s-1)}
- \frac{U}{\nu} ) + U_{X} \right )dT\; , \label{oneform1} \\   
\noalign{\medskip} \noalign{\medskip}\noalign{\medskip}
{\cal T}^{\ast}\omega^{2} & = & 
-\frac{1}{3}\,{\frac {\left (-3\,\epsilon\,U{\nu}^{2
}-\epsilon+\epsilon\,{\nu}^{5/2}+3\,s{\nu}^{5/2}+3\,{\nu}^{4}U_{XX}
\right )}{\nu^{7/2}}} dX                                                     \label{oneform2}  \\
 & + &  \frac{1}{9}\,{\frac{1}{{\nu}^{11/2} }}
       \left( 6{\nu}^{4}(\nu^{5/2}-1)U_{XX} - 9\,\epsilon\,\nu^{4}{U}^{2}
         +9\,{\nu}^{6}U_{XX}U - 9\,U_{X}\frac{\sqrt {\epsilon+1-{s}^{2}}}{1-s} {\nu}^{11/2}  
                                                                    \right.         \nonumber \\
& + & \left. 3\nu^{2}[\epsilon + \nu^{5/2}(3s+\epsilon\frac{s+2}{1-s})]U  
 -     \nu^{5/2}(6s - \epsilon\frac{4s-1}{1-s}) + 2\epsilon -
         \frac{1+2s}{1-s}(3s+\epsilon){\nu}^{5}\right)dT ,   \nonumber \\   
\noalign{\medskip}\noalign{\medskip}\noalign{\medskip}
{\cal T}^{\ast}\omega^{3} & = &
 \left( U_{XX}\sqrt {\nu}-{\frac {\epsilon\,U}{{\nu}^{3/2}}}-\frac{1}{3}\,
        {\frac {\epsilon}{{\nu}^{7/2}}}+\frac{1}{3}\,{\frac {\epsilon}{\nu}}+\frac{1}{\nu} 
\right) dX
+ \frac{1}{s-1} [\, \sqrt{\nu}(1-s)U_{XX}U                             \nonumber \\
 & + & 
\frac{1}{3}[ (1-s)(\frac{\epsilon}{\nu^{7/2}} + \frac{3}{\nu}) + (s+2)\frac{\epsilon}{\nu} ]\,U 
- {\frac{\epsilon}{{\nu}^{3/2}}}(1-s)U^{2} - U_{X}\sqrt{\epsilon+1-{s}^{2}}          \nonumber \\
 & + & 
\frac{2}{3}(1-s)(\frac{\epsilon}{3\nu^{11/2}} - \frac{1}{\nu^{3}}) +
{\frac{\epsilon}{9{\nu}^{3}}}(4s-1)-\frac{1}{3\sqrt{\nu}}(\frac{\epsilon}{3}+1)(1+2s)\nonumber \\
 & + & 
\frac{2}{3}(\nu-\frac{1}{\nu^{3/2}})(1-s)U_{XX} \,]\,dT \; .  \; \; \; \;   \label{oneform3}
\end{eqnarray}

\begin{cor}
The nonlinear equation $(\ref{gen-eqn})$ is geometrically integrable.
\end{cor}

\subsection{Pseudo-potentials}

The one--forms (\ref{oneform1})--(\ref{oneform3}) can be now used to compute the quadratic 
pseudo-potential (\ref{sgamma2}) associated with equation (\ref{gen-eqn}). The resulting 
formulae are very involved, but they can be simplified as follows. After writing down 
(\ref{sgamma2}) with the help of the one--forms $T^{\ast}\omega^{i}$, one applies the 
transformation 
\[
\hat{\Gamma} \mapsto \hat{\gamma} \sqrt{\nu} + \frac{\sqrt{\epsilon +1-s^{2}}}{1-s} \; ,
\]
and changes the parameter $s$ by setting
\[
s-1 = \frac{\sqrt{\nu}}{\lambda} \; .
\] 

\begin{theorem} \label{quadratic}
Equation $(\ref{gen-eqn})$ admits a quadratic pseudo-potential $\hat{\gamma}(X,T)$ determined by
\begin{equation}
{\frac {\partial }{\partial X}}\hat{\gamma} =-\frac{1}{2}\,{\frac {\hat{\gamma}^{2}}{\lambda}}
-{\frac {\epsilon\,U}{{\nu}^{2}}}+\frac{1}{2}\,{\frac {\lambda\,\epsilon}{{\nu}^{2}}}+
\frac{1}{3}\,{\frac {\epsilon}{{\nu}^{3/2}}}+U_{XX}-\frac{1}{3}\,{\frac {\epsilon}{{\nu}^{4}}}
                                                                                  \label{xpart3}
\end{equation} 
and
\begin{eqnarray}
{\frac {\partial }{\partial T}}\hat{\gamma}  & = & 
\left (\frac{1}{2}\,{\frac {U}{\lambda}}- \frac{1}{3}\,
{\frac {1}{{\nu}^{2}\lambda}}+\frac{1}{2}+ \frac{1}{3}\,{\frac {
\sqrt {\nu}}{\lambda}}\right )\,\hat{\gamma}^{2}
-U_{X}\hat{\gamma} - \frac{2}{3}\,U_{XX}\sqrt {\nu}                            \nonumber \\
 &  & - \, \frac{2}{9}\,{\frac {\epsilon}{{\nu}^{6}
}}-U_{XX}U+{\frac {\epsilon\,{U}^{2}}{{\nu}^{2}}}
-\frac{2}{3}\,{\frac {\lambda\,\epsilon}{{\nu}^{3/2}}}+ \frac{2}{3}\,{\frac {\lambda\,
\epsilon}{{\nu}^{4}}}+\frac{1}{2}\,{\frac {\lambda\,\epsilon\,U}{{\nu}
^{2}}}-\frac{1}{2}\,{\frac {{\lambda}^{2}\epsilon}{{\nu}^{2}}} \nonumber \\
 &  & - \; \frac{2}{9}\,{\frac {
\epsilon}{\nu}}-\frac{1}{3}\,{\frac {\epsilon\,U}{{\nu}^{4}}}+\frac{4}{9}\,{
\frac {\epsilon}{{\nu}^{7/2}}}+\frac{1}{3}\,{\frac {\epsilon\,U}{{\nu}
^{3/2}}}+\frac{2}{3}\,{\frac {U_{XX}}{{\nu}^{2}}}                  \; ,   \label{tpart3}
\end{eqnarray}
in which $\lambda \neq 0$ is a real parameter. 
\end{theorem}

Using the pseudo-potential $\hat{\gamma}$ one can simplify the linear problem associated with 
Equation (\ref{gen-eqn}) which follows from (\ref{linear0}) and the one--forms 
${\cal T}^{\ast}\omega^{i}$. Applying Propositions 2 and 4 one finds the following result:

\begin{proposition}
Equation $(\ref{gen-eqn})$, and therefore 
the CH, HS, and KdV equations, is the integrability condition of the one--parameter family
of linear problems $d \psi = (X dx + T dt)\psi$, in which the matrices $X$ and $T$ are given by 
\begin{equation}
X=\left [\begin {array}{cc} 
0 & \displaystyle -\frac{1}{2}\,{\lambda}^{-1}                      \\
\noalign{\medskip} \displaystyle 
\frac{1}{3}\,{\frac {\epsilon}{{\nu}^{4}}}-\frac{1}{2}\,{\frac {\lambda\,\epsilon}
{{\nu}^{2}}}-\frac{1}{3}\,{\frac {\epsilon}{{\nu}^{3/2}}}-U_{XX}+
{\frac {\epsilon\,U}{{\nu}^{2}}} & 0 \end {array}\right ]     \label{equis}
\end{equation}
and
\begin{equation}
T=\left [\begin {array}{cc} 
\displaystyle 
\frac{1}{2}\,U_{X} & 
\displaystyle 
\frac{1}{2}\,{\frac {U}{\lambda}}-\frac{1}{3}\,{\frac {1}{{\nu}^{2}\lambda}}+\frac{1}{2}
+\frac{1}{3}\,{\frac {\sqrt {\nu}}{\lambda}}                                               \\
\noalign{\medskip} \noalign{\medskip} \noalign{\medskip}
\displaystyle 
\frac{2}{3}\,U_{XX}\sqrt{\nu}-\frac{1}{2}{\frac {\lambda\,\epsilon\,U}{{\nu}^{2}}}
-\frac{4}{9}{\frac {\epsilon}{{\nu}^{7/2}}}+ \frac{2}{9}{\frac {\epsilon}{\nu}} 
-\frac{1}{3}{\frac {\epsilon\,U}{{\nu}^{3/2}}} & \\
\noalign{\medskip} \displaystyle
+\frac{2}{3}{\frac {\lambda\,\epsilon}{{\nu}^{3/2}}}
- \frac{2}{3}{\frac {\lambda\,\epsilon}{{\nu}^{4}}}+ 
\frac{2}{9}{\frac {\epsilon}{{\nu}^{6}}} &                                   \\
\noalign{\medskip} \displaystyle
+\frac{1}{2}{\frac {{\lambda}^{2}\epsilon}{{\nu}^{2}}}
- \frac{2}{3}{\frac {U_{XX}}{{\nu}^{2}}}
+U_{XX}U-{\frac {\epsilon\,{U}^{2}}{{\nu}^{2}}}
+ \frac{1}{3}{\frac {\epsilon\,U}{{\nu}^{4}}} & 
\displaystyle -\frac{1}{2}\,U_{X}                                      
     \end {array}\right ]          \; .                          \label{te}
\end{equation}
\end{proposition}

\vspace{8pt}

The linear problem $d \psi = (X dx + T dt)\psi$, with $X,T$ given by (\ref{equis}) and 
(\ref{te}), can be used to find an associated linear problem which is not singular at the KdV 
limit $\nu \rightarrow 0$. Applying the gauge transformation
\[
X_{g} = A X A^{-1} + A_{x}A^{-1} \; , \; \; \; \; \; T_{g} = A T A^{-1} + A_{t}A^{-1} \; ,
\]
in which 
\[
A = \left [\begin {array}{cc} 0&\nu\\\noalign{\medskip}-{\nu}^{-1}&0
\end {array}\right ] \; ,
\]
and changing the parameter $\lambda$ to $\zeta$ by means of 
\begin{equation}
\lambda= (2/3)\,{\nu}^{-2} + (2/3)\,\zeta \; ,    \label{newpar}
\end{equation}
one finds that Equation (\ref{gen-eqn}) is the integrability condition of the linear problem 
$d \psi = (X_{g} dx + T_{g} dt)\psi$ with
\begin{equation}
X_g = \left [\begin {array}{cc} 
      0&(1/3)\,\epsilon\,\zeta+ (1/3)\,
      \sqrt {\nu}\epsilon+{\nu}^{2}U_{XX}-\epsilon\,U \\
      \noalign{\medskip}
      (3/4)\,\left (1+\zeta\,{\nu}^{2}\right )^{-1}&0
             \end {array}
     \right ]
\end{equation}
and
\begin{equation}
T_g = \left [\begin {array}{cc} 
     - \displaystyle \frac{1}{2}\,U_{X} & 
     -\frac{2}{3}\,{\nu}^{5/2} U_{XX}+\frac{1}{3}\,\epsilon\,U\zeta-\frac{2}{9}\,\nu\,\epsilon \\
     & 
       +\frac{1}{3}\,\sqrt{\nu}\epsilon\,U-\frac{4}{9}\,\sqrt {\nu}\epsilon\,\zeta
       -\frac{2}{9}\,\epsilon\,{\zeta}^{2} \\
     & 
       +\frac{2}{3}\,U_{XX}-{\nu}^{2}U_{XX}U+\epsilon\,{U}^{2}\\
       \noalign{\medskip} \displaystyle 
       - {\frac{(3/4)\,U +(1/2)\,\zeta + (1/2)\,\sqrt {\nu}}{1+\zeta\,{\nu}^{2}}}
    & \displaystyle \frac{1}{2} \,U_{X}
     \end {array}
\right ] \; .
\end{equation}

\begin{example}
If $\nu =1$, Equation (\ref{newpar}) implies that the matrices $X_g$ and $T_g$ become
\begin{equation}
X_g = \frac{1}{2}
\left [
\begin {array}{cc}
  0 &  \epsilon\,\lambda + 2\,m  \\
  \lambda^{-1} & 0
\end {array}
\right ] \; , \; \; \; \; 
T_g = \frac{1}{2}
\left [
\begin {array}{cc}
\displaystyle
- {u_{x}} & - 2 u  \, m + \epsilon\,\lambda u  - \epsilon\,\lambda^{2} \\
- 1 - u \lambda^{-1} & u_{x}
\end {array}
\right ]   \; ,                       \label{l4}
\end{equation}
in which $m = u_{xx} - \epsilon u$, and one recovers the associated linear problems for the 
CH and HS equations derived in \cite{R5}.
\end{example}

\begin{cor}
{\em (a)} The nonlinear equation
\begin{equation}
-2\,{\nu}^{2}U_{X}U_{XX}+3\,\epsilon\,U_{X}\,U - {\nu}^{2}U_{XXX}\,U + 
\frac{2}{3}\,U_{XXX}-\frac{2}{3}\,{\nu}^{5/2}U_{XXX}+\epsilon\,U_{T}-{\nu}^{2}U_{XXT} = 0 
										\label{gen-eqn1}
\end{equation}
describes pseudo-spherical surfaces with associated one--forms $\omega^{i} = f_{i 1}
dx + f_{i 2}dt$ in which
\begin{eqnarray}
f_{11} & = & (1/3)\,\epsilon\,\zeta+ (1/3)\,\sqrt {\nu}\epsilon+{\nu}^{2}U_{XX}-\epsilon\,U +
             (3/4)\,\left (1+\zeta\,{\nu}^{2}\right )^{-1}\; , \\
f_{12} & = & (2/3)\left(1-\nu^{5/2}\right)U_{XX} + (\epsilon/3)\left(\zeta+\sqrt{\nu}\right)U
             +\epsilon U^{2} - \nu^{2}U U_{XX} \nonumber \\
       &   & - (2/9)\epsilon (\zeta^{2} + 2 \sqrt{\nu}\zeta + \nu) - 
             \frac{1}{2(1+\zeta \nu^{2})}\left( (3/2)U + \zeta + \sqrt{\nu} \right) \; , \\
f_{21} & = & 0 \; , \\
f_{22} & = & - U_{X} \; , \\
f_{31} & = & -(1/3)\,\epsilon\,\zeta - (1/3)\,\sqrt {\nu}\epsilon - {\nu}^{2}U_{XX}+\epsilon\,U +
             (3/4)\,\left (1+\zeta\,{\nu}^{2}\right )^{-1} \; , \\
f_{32} & = & -(2/3)\left(1-\nu^{5/2}\right)U_{XX} - (\epsilon/3)\left(\zeta+\sqrt{\nu}\right)U
             -\epsilon U^{2} + \nu^{2}U U_{XX} \nonumber \\
       &   & + (2/9)\epsilon (\zeta^{2} + 2 \sqrt{\nu}\zeta + \nu) - 
              \frac{1}{2(1+\zeta \nu^{2})}\left( (3/2)U + \zeta + \sqrt{\nu} \right) \; , 
\end{eqnarray}
and $\zeta$ is an arbitrary parameter.

{\em (b)} Equation $(\ref{gen-eqn1})$ admits the quadratic pseudo-potential $\gamma(X,T)$
determined by the Pfaffian system 
\begin{eqnarray}
- \gamma_{X} & = & \frac{3}{4(1+\zeta\nu^{2})} \gamma^{2} - \left( \nu^{2} U_{XX} - \epsilon U
                   + \frac{1}{3} \epsilon (\zeta + \sqrt{\nu}) \right) \; , 
                   \label{miurax} \\
- \gamma_{T} & = &  \frac{1}{2(1+\zeta\nu^{2})} \left( - \frac{3}{2}U - \zeta - \sqrt{\nu}\right)
                    \gamma^{2} + U_{X} \, \gamma                       \label{miurat}    \\
             & + & \left( U(\nu^{2} U_{XX}-\epsilon U) + \frac{2}{3}(\nu^{5/2}-1)U_{XX}
                  - \frac{1}{3} \epsilon(\zeta+\sqrt{\nu})U
                  + \frac{2}{9} \epsilon (\zeta^{2} + 2 \sqrt{\nu} \zeta 
                  + \nu)\right) \; .                          \nonumber 
\end{eqnarray}

\end{cor}

\begin{example}  \label{cl-ej}
Taking $\nu = 0$ and $\epsilon = 1$ in (\ref{miurax}) gives 
\[
- \gamma_{X} = \frac{3}{4}\gamma^{2} + U - \frac{1}{3}\zeta \; ,
\]
and one recovers the usual Miura transformation for the KdV equation. On the other hand, taking 
$\nu =1$ and $\lambda = (2/3)(1+\zeta)$ in (\ref{miurax}) and (\ref{miurat}) yields
\begin{eqnarray}
U_{XX} - \epsilon U & = & \gamma_{X} + \frac{\gamma^{2}}{2\lambda}-\frac{\epsilon}{2}\lambda \;,
									\label{miurax1}      \\
- \gamma_{T} & = & - \frac{1}{2}(\frac{U}{\lambda} + 1)\gamma^{2} + U_{X}\gamma +
                   (\, U(U_{XX}-\epsilon U) - \frac{1}{2}\epsilon U \lambda + \frac{1}{2}\epsilon
                      \lambda^{2} \,)        \; ,                              \label{miurat1}
\end{eqnarray}
and substitution of (\ref{miurax1}) into (\ref{miurat1}) implies that the Camassa--Holm equation 
$(\ref{ch})$ and the Hunter--Saxton equation $(\ref{hs})$ possess the parameter--dependent 
conservation law
\begin{equation}
\gamma_{T}  =  \lambda \, (\,u_{X} - \gamma - \frac{1}{\lambda} u\gamma\,)_{X} \; . \label{cl3}
\end{equation}
As in the KdV case, one can use (\ref{miurax1}) and (\ref{cl3}) to construct conservation laws 
for the CH and HS equations \cite{CH,s-asso,HZ,R4,R5}. Setting $\gamma = \sum_{n=1}^{\infty} 
\gamma_{n} \lambda^{n/2}$ yields the conserved densities 
\begin{eqnarray}
\gamma_{1} & = & \sqrt{2}\,\sqrt{m} \;, \; \; \; \gamma_{2} \, = \,  - \frac{1}{2}\,\ln (m)_{X}  
\; , \; \; \; 
\gamma_{3} \, = \, \frac{1}{2\sqrt{2}\,\sqrt{m}}\,[ \epsilon - \frac{m_{X}^{2}}{4\,m^{2}} + 
                                                         \ln (m)_{XX}]\, , \;\; \label{cd-1} \\
\gamma_{n+1} & = & - \frac{1}{\gamma_{1}} \, \gamma_{n,X} - \frac{1}{2\gamma_{1}} \,
\sum_{j=2}^{n} \gamma_{j} \, \gamma_{n+2-j} \; , \; \; \; \; n \geq 3 \; ,         \label{cd-4}
\end{eqnarray}  
in which $m = u_{xx} - \epsilon u$, while the expansion $\gamma = \epsilon\,\lambda + 
\sum_{n=0}^{\infty} \gamma_{n} \lambda^{-n}$ implies 
\begin{equation}
\gamma_{0,X} + \epsilon\,\gamma_{0} = m \; , \; \; \; \; \; 
\gamma_{n,X} + \epsilon\,\gamma_{n} = - (1/2) \sum_{j=0}^{n-1} \gamma_{j} \, \gamma_{n-1-j} \; , 
                                 \; \; \; \;         n \geq 1 \; .                    \label{cd2}
\end{equation}                                                                     
In the CH case, (\ref{cd2}) allows one to recover the familiar local conserved densities $u$, 
$u_{X}^{2} + u ^{2}$, and $u u_{X}^{2} + u ^{3}$, \cite{CH}, and a sequence of nonlocal 
conservation laws. 
\end{example}

In view of Examples \ref{ej} and \ref{cl-ej}, it is natural to postulate Equation (\ref{miurax1})
as the analog of the Miura transformation for the CH and HS equations, and (\ref{cl3}) as the 
corresponding ``modified'' equation. Note that, in contradistinction with the 
KdV case, the modified CH and HS equations are nonlocal equations for $\gamma$. 

\begin{remark}
The question whether there exists a modified CH equation has been asked by J. Schiff 
in \cite{s-dual}. Earlier contributions to this problem have been made by Fokas \cite{fo}, 
Fuchssteiner \cite{fu}, Schiff \cite{s-loop}, and Camassa and Zenchuk \cite{cz}. The reader 
is referred to \cite{R4} for a discussion on the relation between these works and the modified 
equation proposed here.
\end{remark}

\subsection{Nonlocal symmetries}

In this subsection is shown that one can find a nonlocal symmetry of Equation (\ref{gen-eqn1}) 
depending on the pseudo-potential $\gamma (X,T)$ given by (\ref{miurax}), (\ref{miurat}). To 
begin with, note that substitution of (\ref{miurax}) into (\ref{miurat}) yields the conservation
law
\begin{equation}
\gamma_{T} = \left[ \frac{2}{3}(\zeta \nu^{2} + 1)U_{X} - \frac{2}{3}(\zeta + \sqrt{\nu})\gamma
               - \gamma U \right]_{X} \; .                                            \label{cl4}
\end{equation}
In analogy with the Camassa-Holm case \cite{R4}, one has the following result:
\begin{theorem}                                                               \label{shadow-sym}
Set $m = \nu^{2} U_{XX} - \epsilon U$, and let $\gamma$ and $\delta$ be defined by the equations 
\begin{equation}
\gamma_{X} = \frac{- 3}{4(1+\zeta\nu^{2})} \gamma^{2} 
              + \left( m + \frac{1}{3} \epsilon (\zeta + \sqrt{\nu}) \right)\, , \; \; \; 
\gamma_{T}  =  \left[ \frac{2}{3}(\zeta \nu^{2} + 1)U_{X} - \frac{2}{3}(\zeta + \sqrt{\nu})\gamma
               - \gamma U \right]_{X}  \, ,                                     \label{g-shadow} 
\end{equation}
and
\begin{equation}
\delta_{X} = \gamma \, ,  \; \; \; \; \; \; \; \; \; \; \; \; \; 
\delta_{T} = \frac{2}{3}(\zeta \nu^{2}+1)U_{X} - \frac{2}{3}(\zeta + \sqrt{\nu})\gamma 
                                               - U \gamma \; ,                 \label{d-shadow} 
\end{equation}
which are compatible on solutions of $(\ref{gen-eqn1})$. The nonlocal vector field
\begin{equation}
V = \gamma \, \exp ( \,(3/2) \frac{\delta}{1+\zeta\nu^{2}}\, ) \,\frac{\partial}{\partial U} 
									\label{vec-shadow}
\end{equation}
determines a shadow of a nonlocal symmetry for the nonlinear equation $(\ref{gen-eqn1})$.
\end{theorem}
\begin{proof}
Define a covering $\overline{S}$ of the equation manifold of (\ref{gen-eqn1}) as follows.
Locally, $\overline{S}$ is equipped with coordinates $(X,T,U,\dots,U_{X^{p}T^{q}},\dots,\gamma,
\delta)$, in which $p > 0$, and the total derivatives $\overline{D}_{X}$ and $\overline{D}_{T}$ 
are given by 
\begin{eqnarray*}
\overline{D}_{X} & = & D_{X} + \left( \frac{- 3}{4(1+\zeta\nu^{2})} \gamma^{2} 
                + m + \frac{1}{3} \epsilon (\zeta + \sqrt{\nu}) \right)
                \frac{\partial}{\partial \gamma} + \gamma \frac{\partial}{\partial \delta}\; ,\\ 
\overline{D}_{T} & = & D_{T} + 
                    \left( \frac{1}{2(1+\zeta\nu^{2})}\left(\frac{3}{2}U+\zeta +\sqrt{\nu}\right)
                      \gamma^{2} - U_{X} \, \gamma - U(\nu^{2} U_{XX}-\epsilon U) \right. \\
         &  & \left. \; \; \; \; \; \; 
              - \; 
              \frac{2}{3}(\nu^{5/2}-1)U_{XX} + \frac{1}{3} \epsilon(\zeta+\sqrt{\nu})U
              - \frac{2}{9} \epsilon (\zeta^{2} + 2 \sqrt{\nu} \zeta  + \nu)\right)
                                                          \, \frac{\partial}{\partial \gamma} \\
         &  & \; \; \; \; \; \; 
              + \; 
              \left( \frac{2}{3}(\zeta \nu^{2}+1)U_{X} - \frac{2}{3}(\zeta + \sqrt{\nu})\gamma
                                        - U \gamma \right) \, \frac{\partial}{\partial \delta}\; .
\end{eqnarray*}

Now, the vector field $V$ determines the shadow of a nonlocal symmetry if the function
\begin{equation}
G = \gamma \, \exp ( \,(3/2) \frac{\delta}{1+\zeta\nu^{2}}\, )      \label{gshadow}
\end{equation}
satisfies the equation
\[
\overline{D}_{T} G = \sum_{i=0}^{3} \frac{\partial F}{\partial u_{X^{i}}}\overline{D}_{X}^{i}(G)
                 + \frac{\partial F}{\partial u_{XXT}}\overline{D}_{X}^{2}\overline{D}_{T}(G)    
\]
identically, in which $F$ is the right hand side of Equation (\ref{gen-eqn1}) when written as 
$U_{T} = F$. Checking that this is so is a long but straightforward computation. It can be done 
using the MAPLE package VESSIOT developed by I. Anderson and his coworkers, see \cite{ian}.
\end{proof}

The next problem to be considered is the extension of the shadow (\ref{vec-shadow}) to a bona 
fide nonlocal symmetry. For this, one studies the variations of the functions $\gamma$ and 
$\delta$ induced by the infinitesimal deformation $U \mapsto U + \tau G$, in which $G$ is given
by (\ref{gshadow}).

Note that Equation (\ref{gen-eqn1}) can be written as a system of equations for 
two variables, $m$ and $U$, as follows:
\begin{equation}
m = \nu^{2}U_{XX} - \epsilon U \, , \; \; \; \; \; \; \; \; 
m_{T} = - m_{X}U - 2m U_{X} + \frac{2}{3}(1 - \nu^{5/2})U_{XXX} \; .   \label{gen-sys}
\end{equation}

\begin{theorem}  \label{sy}
Let $\gamma$, $\delta$ and $\beta$ be defined by the equations 
\begin{eqnarray}
\gamma_{X}  & = & - \frac{3}{4(1+\zeta\nu^{2})} \gamma^{2} 
                  + \left( m + \frac{1}{3}\epsilon (\zeta + \sqrt{\nu}) \right)\, , \label{g-x}\\
\gamma_{T}  & = & \left[ \frac{2}{3}(\zeta \nu^{2} + 1)U_{X}-\frac{2}{3}(\zeta + \sqrt{\nu})\gamma
                  - \gamma U \right]_{X}  \; ,                                     \label{g-t} \\
\delta_{X} & = & \gamma \, ,                                                     \label{d-x} \\
\delta_{T} & = & \frac{2}{3}(\zeta \nu^{2}+1)U_{X} - \frac{2}{3}(\zeta + \sqrt{\nu})\gamma 
                                  - U \gamma \; ,                               \label{d-t}  \\
\beta_{X} & = & \left[\nu^{2} m + (1/3)\epsilon (\nu^{5/2}-1)\right]\exp 
                ( \,(3/2)\frac{\delta}{1+\zeta\nu^{2}}\, ) \; ,               \label{b-x}     \\
\beta_{T} & = & \left[ - \frac{1}{3}(\nu^{5/2}-1)(2m+\epsilon U) 
               - \frac{1}{2}\,\gamma^{2} + \frac{2}{9} \, \epsilon \, (2 \zeta + \zeta^{2}\nu^{2}
                - \nu^{3} + 2 \sqrt{\nu})    \right.                          \nonumber \\
          &   & \left.\;\,\vphantom{\frac{1}{3}} - \nu^{2} U \,m \right]
                \exp ( \,(3/2) \frac{\delta}{1+\zeta\nu^{2}}\, )   \; ,          \label{b-t}
\end{eqnarray}
which are compatible on solutions of $(\ref{gen-sys})$. The system 
of equations $(\ref{gen-sys})$--$(\ref{b-t})$ possesses the classical 
symmetry
\begin{eqnarray}
W & = & \gamma \, \exp ( \,(3/2) \frac{\delta}{1+\zeta\nu^{2}}\, ) \frac{\partial}{\partial U} 
									\nonumber \\
  &   & + \, \left[\nu^{2}m_{X}+\frac{3\nu^{2}\gamma}{1+\zeta\nu^{2}}m
        +\gamma\epsilon\frac{\nu^{5/2}-1}{1+\zeta\nu^{2}}\right]
         \exp( \,(3/2) \frac{\delta}{1+\zeta\nu^{2}}\, ) \frac{\partial}{\partial m} \nonumber\\
  &  &  + \, \left[ \nu^{2} m + \frac{1}{3} \epsilon (\nu^{5/2}-1)\right]\,
      \exp( \,(3/2) \frac{\delta}{1+\zeta\nu^{2}}\, ) \frac{\partial}{\partial\gamma} \nonumber\\
  &  &  + \beta \, \frac{\partial}{\partial\delta}                                  \nonumber \\ 
  &  &  + \left(\nu^{2}[\nu^{2}m + \frac{1}{3}\epsilon(\nu^{5/2}-1)]\exp(\,3\frac{\delta}
          {1+\zeta\nu^{2}}\,)
        + \frac{3}{4(1+\zeta\nu^{2})}\,\beta^{2}\right) \frac{\partial}{\partial\beta} \; .
                                                                                      \label{nls}
\end{eqnarray}
\end{theorem}

As with Theorem \ref{shadow-sym}, Theorem \ref{sy} can be verified using the MAPLE package 
VESSIOT \cite{ian}. In terms of the theory of coverings, one has

\begin{cor}   \label{nonloc}
The vector field $(\ref{nls})$ determines a nonlocal symmetry of the system of equations 
$(\ref{gen-sys})$.
\end{cor}
\begin{proof}
Define a covering $\overline{S}$ of the equation manifold of Equation (\ref{gen-sys}) as follows:
locally, $\overline{S}$ is equipped with coordinates $(X,T,U,U_{X},U_{T},\dots ,U_{X^{2p+1}T^{q}}
, \dots , m,\dots , m_{X^{r}}, \dots , \gamma , \delta , \beta)$, and the total derivatives
$\overline{D}_{X}$ and $\overline{D}_{T}$ are given by
\begin{eqnarray*}
\overline{D}_{X} & = & D_{X} + \left( \frac{- 3}{4(1+\zeta\nu^{2})} \gamma^{2} 
                       + m + \frac{1}{3} \epsilon (\zeta + \sqrt{\nu}) \right)
                \frac{\partial}{\partial \gamma} + \gamma \frac{\partial}{\partial \delta} \\
               &    & \; \; \; \; \; \; \, + \left[\nu^{2} m + (1/3)\epsilon (\nu^{5/2}-1)\right]
             \exp( \,(3/2)\frac{\delta}{1+\zeta\nu^{2}}\, )\frac{\partial}{\partial\beta}\; , \\ 
\overline{D}_{T} & = & D_{T} + 
                    \left( \frac{1}{2(1+\zeta\nu^{2})}\left(\frac{3}{2}U+\zeta +\sqrt{\nu}\right)
                      \gamma^{2} - U_{X} \, \gamma - U(\nu^{2} U_{XX}-\epsilon U) \right. \\
         &  & \left. \; \; \; \; \; \; 
              - \; 
              \frac{2}{3}(\nu^{5/2}-1)U_{XX} + \frac{1}{3} \epsilon(\zeta+\sqrt{\nu})U
              - \frac{2}{9} \epsilon (\zeta^{2} + 2 \sqrt{\nu} \zeta  + \nu)\right)
                                                          \, \frac{\partial}{\partial \gamma} \\
         &  & \; \; \; \; \; \; 
              + \; 
              \left( \frac{2}{3}(\zeta \nu^{2}+1)U_{X} - \frac{2}{3}(\zeta + \sqrt{\nu})\gamma
                                      - U \gamma \right) \, \frac{\partial}{\partial \delta}\; \\
         &  &  \; \; \; \; \; \;
              + \left[ - \frac{1}{3}(\nu^{5/2}-1)(2m+\epsilon U) 
               - \frac{1}{2}\,\gamma^{2} + \frac{2}{9} \, \epsilon \, (2 \zeta + \zeta^{2}\nu^{2}
                - \nu^{3} + 2 \sqrt{\nu}) - \nu^{2} U \,m \right]\times \\
         &  &  \; \; \; \; \; \; \; \; \; \; 
               \exp ( \,(3/2) \frac{\delta}{1+\zeta\nu^{2}}\, )\frac{\partial}{\partial\beta}\; .
\end{eqnarray*}
One now sets
\begin{eqnarray*}
G_{1} & = & \gamma \, \exp ( \,(3/2) \frac{\delta}{1+\zeta\nu^{2}}\, ) \; , \\
G_{2} & = & \left[\nu^{2}m_{X}+\frac{3\nu^{2}\gamma}{1+\zeta\nu^{2}}m
            +\gamma\epsilon\frac{\nu^{5/2}-1}{1+\zeta\nu^{2}}\right]
            \exp( \,(3/2) \frac{\delta}{1+\zeta\nu^{2}}\, ) \; \\
I_{\gamma} & = & \left[ \nu^{2} m + \frac{1}{3} \epsilon (\nu^{5/2}-1)\right]\,
                 \exp( \,(3/2) \frac{\delta}{1+\zeta\nu^{2}}\, ) \; , \\
I_{\delta} & = & \beta \; , \\
I_{\beta} & = &  \nu^{2}[\nu^{2}m + \frac{1}{3}\epsilon(\nu^{5/2}-1)]\exp(\,3\frac{\delta}
          {1+\zeta\nu^{2}}\,)  + \frac{3}{4(1+\zeta\nu^{2})}\,\beta^{2} \; .
\end{eqnarray*}
Then, the vector field
\[
\overline{D}_{\tau} = \sum_{p,q}\overline{D}_{X}^{\,p}\overline{D}_{T}^{\,q}(G_{1})
                      \frac{\partial}{\partial U_{X^{p}T^{q}}} + 
                      \sum_{i \geq 0}\overline{D}_{X}^{\,i}\frac{\partial}{\partial m_{X^{i}}} +
                      I_{\gamma}\frac{\partial}{\partial \gamma} +
                      I_{\delta}\frac{\partial}{\partial \delta} +
                      I_{\beta}\frac{\partial}{\partial \beta}
\]
is a nonlocal symmetry of the system of equations (\ref{gen-sys}). In fact, the fist equation
of (\ref{shadow2}) becomes
\[
\left(
\begin{array}{cc}
-\nu^{2}\overline{D}_{X}^{\,2} + \epsilon & 1 \\
\frac{2}{3}(1-\nu^{5/2})\overline{D}_{X}^{\,3} + 2 m \overline{D}_{X} + m_{X} &
\overline{D}_{T} + U \overline{D}_{X} + 2 U_{X}
\end{array}
\right)
\left(
\begin{array}{c}
G_{1} \\ G_{2}
\end{array}
\right)
=
\left(
\begin{array}{c}
0 \\ 0
\end{array}
\right) \; , 
\]
and (\ref{g-x})--(\ref{b-t}) imply that this equation is equivalent to the fact 
that $G_{1}$ and $G_{2}$ satisfy the linearization  of (\ref{gen-sys}). The other equations
of (\ref{shadow2}) hold because the functions $I_{\gamma}$, $I_{\delta}$ and $I_{\beta}$
satisfy the linearizations of Equations (\ref{g-x})--(\ref{b-t}). 
\end{proof}

It is clear from Corollary \ref{nonloc} and its proof that the results of Galas \cite{Ga}, Leo 
{\em et. al.} \cite{llst} and Schiff \cite{s-pre} mentioned in Section 1, can be 
also interpreted in terms of coverings and provide further examples of nonlocal symmetries.

The advantage of the vector field $W$ given by (\ref{nls}) over the shadow $V$ defined in 
(\ref{vec-shadow}) is that one can find the flow of $W$ simply by integrating a first
order system of partial differential equations, and therefore one can obtain a (local) existence 
theorem for solutions of the nonlinear equation (\ref{gen-sys}). Consider the following
first order system in independent variables $\xi$ and $\eta$:
\begin{eqnarray}
\frac{\partial\, x}{\partial\xi} &=& - \nu^{2}{\rm e}^{D (\xi,\eta)} \; ,        \label{sys1}\\
\frac{\partial \, m}{\partial \xi} & = & \frac{1}{1+\zeta\nu^{2}} \left(3\nu^{2} m(\xi,\eta) 
   + \epsilon (\nu^{5/2}-1)\right)\gamma(\xi,\eta)\,{\rm e}^{D(\xi,\eta)} \; ,    \label{sys2} \\
\frac{\partial \, \gamma}{\partial \xi} & = & \left( \frac{3\nu^{2}}{4(1+\zeta\nu^{2})}\,
      \gamma(\xi,\eta)^{2} - \frac{1}{3}\,\epsilon(1+\zeta\nu^{2})\right){\rm e}^{D(\xi,\eta)} 
                                                                                     \,  \; , \\
\frac{\partial \, \delta}{\partial \xi} & = & \beta(\xi,\eta) - \nu^{2}\gamma(\xi,\eta) \, 
                                             {\rm e}^{D (\xi,\eta)}  \; ,   \\
\frac{\partial \, \beta}{\partial \xi} & = & \frac{3}{4(1+\zeta\nu^{2})}\,\beta(\xi,\eta)^{2}\; ,
                                                                                     \label{sys5}
\end{eqnarray}
in which 
\begin{equation}
D(\xi,\eta)=\frac{3\delta(\xi,\eta)}{2(1+\zeta\nu^{2})} \, .    \label{de}
\end{equation}

\begin{proposition} \label{sy2}
The system of equations $(\ref{sys1})$--$(\ref{sys5})$ with initial conditions $\beta_{0} = 
\beta(0,\eta)$, $\gamma_{0} = \gamma(0,\eta)$, $\delta_{0} = \delta(0,\eta)$, $m_{0} = 
m(0,\eta)$, and $X_{0} = X(0,\eta) = \eta$, has the solution
\begin{eqnarray}
X (\xi,\eta) & = & -\nu^2 \int_{0}^{\xi}{\rm e}^{D(z,\eta)}\,dz + \eta \; ,      \label{tra}  \\
\ln \left|{\frac {3\,{\nu}^{2}m(\xi,\eta)+\epsilon\,\left ({\nu}^{5/2}-1\right )
}{3\,{\nu}^{2}m_{0}+\epsilon\,\left ({\nu}^{5/2}-1\right )}}\right| & = &
\frac{3\,{\nu}^{2}}{(1+\zeta\,{\nu}^{2})} \int _{0}^{\xi}\gamma(z,\eta){\rm e}^{D(z,\eta)}\,du 
                                                                                 \label{sol4} \\
\gamma (\xi,\eta) & = & \frac{1}{9}\left({\frac {-4(1+\zeta\,{\nu}^{2})}
                {\left(-3\,\xi\,\beta_{0}+4+4\,\zeta\,{\nu}^{2}\right )\beta_{0}}}
                +\frac{1}{\beta_{0}}\right) \times                                \nonumber \\
                &   & \; \; \; \, \left(4\,\epsilon\left (1+\zeta\,{\nu}^{2}\right )^{2}
               -9\,{\nu}^{2}{\gamma_{0}}^{2}\right) {\rm e}^{D(0,\eta)} + \gamma_{0}\; ,
                                                                                 \label{sol2} \\
\delta (\xi,\eta) & = & \frac{2}{3}(1+\zeta\nu^2)\ln\left|\frac{4(1+\zeta\nu^2)
         {\displaystyle \frac{\partial\gamma}{\partial \xi}}(\xi,\eta)}{3\nu^2\gamma(\xi,\eta)^2 
                      - (4/3)\epsilon(1+\zeta\nu^2)^2}\right| \, , \; \; \; \; \;  \label{sol3}\\
\beta (\xi,\eta) & = & 4\,\frac {(1+\zeta\,{\nu}^{2})\beta_{0}}
                       {-3\,\xi\,\beta_{0}+4+4\,\zeta\,\nu^{2}} \; ,         \label{sol1} 
\end{eqnarray}
in which the functions $D(0,\eta)$ and $D(z,\eta)$ are determined by $(\ref{de})$, the initial
condition $\delta_{0} = \delta(0,\eta)$, and Equations $(\ref{sol2})$ and $(\ref{sol3})$. 
\end{proposition}

Now, Equation (\ref{tra}) determines a transformation $(\xi , \eta ) \mapsto (X, \tau)$, in
which $\tau$ is a parameter along the flow of $W$, given by, say,
\begin{equation}
\tau = \xi \; , \; \; \; \; \; \; \; \; \; \; X = h(\xi , \eta) \; . \label{trans}
\end{equation}
Applying this change of variables to Equations (\ref{sys2})--(\ref{sys5}), and using (\ref{sys1})
and Equations (\ref{g-x}), (\ref{d-x}), and (\ref{b-t}) for $\gamma$, $\delta$, and $\beta$, 
one sees that formulae (\ref{sol4})--(\ref{sol1}) provide solutions for the flow equations
\begin{eqnarray}
\frac{\partial \, m}{\partial \tau} & = & \left[\nu^{2}m_{X} +
\frac{3\nu^{2}\gamma}{1+\zeta\nu^{2}}m +\gamma\epsilon\frac{\nu^{5/2}-1}{1+\zeta\nu^{2}}\right]
         \exp( \,(3/2) \frac{\delta}{1+\zeta\nu^{2}}\, ) \; ,   \\
\frac{\partial \, \gamma}{\partial \tau} & = & \left[\nu^{2} m + 
\frac{1}{3}\epsilon (\nu^{5/2}-1)\right]\,\exp( \,(3/2) \frac{\delta}{1+\zeta\nu^{2}}\, )  \; , \\
\frac{\partial \, \delta}{\partial \tau} & = & \beta  \; ,      \\
\frac{\partial \, \beta}{\partial \tau} & = & \nu^{2}
          \left[\nu^{2}m+\frac{1}{3}\epsilon(\nu^{5/2}-1)\right]
          \exp(\,3\frac{\delta}{1+\zeta\nu^{2}}\,) + \frac{3}{4(1+\zeta\nu^{2})}\,\beta^{2} \; ,
\end{eqnarray}
which one obtains from the formula for $W$ in Theorem \ref{sy}. Thus, finding a 
two--parameters (the ``flow'' parameter $\tau$ and the ``spectral'' parameter $\zeta$) family 
of solutions to the nonlinear equation  (\ref{gen-sys}) amounts to solving {\em one} simple 
equation. More exactly, one has,

\begin{cor} \label{newsol}
Let $U(X,T)$ be a solution of Equation $(\ref{gen-sys})$. Then, the solution $U(X,T,\tau)$ to 
the initial value problem
\begin{eqnarray}
\frac{\partial \, U}{\partial \tau} & = & 
                 \gamma (X,T,\tau)\exp(\,(3/2)\frac{\delta(X,T,\tau)}{1+\zeta\nu^{2}}\, )  \, ,\\
U(X,T,0) & = & U(X,T),
\end{eqnarray}
in which $\gamma (X, T, \tau)$ and $\delta(X,T,\tau)$ are determined by $(\ref{sol2})$, 
$(\ref{sol3})$, and $(\ref{trans})$, is a two--parameters family of solutions to Equation 
$(\ref{gen-sys})$.
\end{cor}

This paper ends with two elementary examples. 

\begin{example}
In the Camassa--Holm case, $\nu = 1$, $\epsilon =1$, $\lambda = (2/3)(1+\zeta)$, the first 
order system (\ref{sys1})--(\ref{sys5}) becomes 
\begin{eqnarray}
\frac{\partial X}{\partial\xi} \; = \; -{\rm e}^{\delta/\lambda}\, , \; \; &  & \; \; 
\frac{\partial m}{\partial\xi} \; = \; \frac{2}{\lambda}\,\gamma\,{\rm e}^{\delta/\lambda}\,m,
 \label{special1} \\
{\frac {\partial\gamma}{\partial\xi}} \; = \; - \frac{1}{2\lambda}\,{\rm e}^{\delta/\lambda}
\left ({\lambda}^{2} - \gamma^{2}\right), \; \;  &  & \; \; 
{\frac {\partial\delta}{\partial\xi}} \; = \; \beta -\gamma\,{\rm e}^{\delta/\lambda}\, , 
\; \; \; \; \; \; \; \; \; 
\frac{\partial\beta}{\partial\xi} \; = \; \frac{1}{2\lambda}\,{\beta^{2}}, \label{special2}
\end{eqnarray}
and the solutions (\ref{tra})--(\ref{sol1}) now read
\begin{eqnarray}
X & = & \eta + \ln \left|\frac{-\xi\,\beta_{0}+2\,\lambda
   +(\gamma_{0}-\lambda)\xi\,{\rm e}^{\delta_{0}/\lambda} }
   {-\xi\,\beta_{0}+2\,\lambda+(\gamma_{0}+\lambda)\xi\,{\rm e}^{\delta_{0}/\lambda} }\right| 
                                                                                \label{new-x} \\
m & = & \frac{m_{0}}{(-\xi\,\beta_{0}+2\,\lambda)^{4}}
   (-\xi\beta_{0}+2\lambda+(\gamma_{0}-\lambda)\xi{\rm e}^{\delta_{0}/\lambda})^{2}
   (-\xi\beta_{0}+2\lambda+(\gamma_{0}+\lambda)\xi{\rm e}^{\delta_{0}/\lambda})^{2} \;\;\;\;\; \\
\gamma & = &  \gamma_{0} + \frac{\xi\,(\gamma_{0}^{2}-\lambda^{2})}{-\xi\,\beta_{0}+2\,\lambda}\,
              {\rm e}^{\delta_{0}/\lambda}            \label{new-gamma}                   \\
\delta & = & \lambda\,\ln \left| \frac{4\,{\lambda}^{2}{\rm e}^{\delta_{0}/\lambda} }
   {(-\xi\,\beta_{0}+2\,\lambda+(\gamma_{0}+\lambda)\xi\,{\rm e}^{\delta_{0}/\lambda})
   (-\xi\,\beta_{0}+2\,\lambda+(\gamma_{0}-\lambda)\xi\,{\rm e}^{\delta_{0}/\lambda})}\right|
  \label{new-delta}  \\
\beta & = &  2\,{\frac {\lambda\,\beta_{0}}{-\xi\,\beta_{0}+2\,\lambda}} \; . \label{new-beta}
\end{eqnarray}
These formulae appear in \cite{R4}. Now consider the Camassa--Holm equation in the form
\begin{equation}
2 U_{\eta}U_{\eta\eta} + U U_{\eta\eta\eta} = U_{T} - U_{\eta\eta T} + 3U_{\eta}U \; ,\label{old}
\end{equation}
so that the ``old'' space variable is $\eta$, and choose an obvious solution of (\ref{old}), say
$U_{0}(\eta, T) = e^{\eta}$. 
The corresponding (pseudo)potentials $\gamma_{0}$, $\delta_{0}$ and $\beta_{0}$, computed by 
means of (\ref{g-x})--(\ref{b-t}), are given by
\[
\gamma_{0} = \lambda \; , \; \; \; \; \; \; \; \; \beta_{0} = c \; , \; \; \; \; \; \; \; \; 
\delta_{0} = \lambda \eta - \lambda^{2} T.
\]
Use these values as initial conditions for the system (\ref{special1}), (\ref{special2}) that is,
take
\[
U_{0}(\eta, T) = e^{\eta} \; , \; \; \; \; m_{0} = 0 \; , \; \; \; \; 
\gamma_{0} = \lambda \; , \; \; \; \; \delta_{0} = \lambda
\eta - \lambda^{2} T \; , \; \; \; \; \beta_{0} = c \; .
\]
The new space variable $X$ is then given by Equation (\ref{new-x}). One finds  
\begin{equation}
X(\xi,\eta, T) = \eta + \ln \left| \frac{\xi c + 2 \lambda}{-\xi c + 2 \lambda + \xi e^{\eta - 
\lambda T} 2\lambda} \right| \; ,     \label{x-eqn}
\end{equation}
while the (pseudo)potentials $\gamma$, $\delta$ and $\beta$ become
\begin{eqnarray}
\gamma (\xi, \eta , T) & = & \lambda \; , \\
\delta (\xi , \eta , T) & = & \lambda \ln \left| \frac{4 \lambda^{2}{\rm e}^{\eta - \lambda T}}
                              { (-\xi c + 2\lambda + 2\lambda\xi{\rm e}^{\eta - \lambda T})
                                (-\xi c + 2\lambda)} \right| \; , \\
\beta (\xi ,\eta, T) & = &  \frac{2\lambda c}{-\xi c + 2 \lambda} \; .
\end{eqnarray}

Now invert Equation (\ref{x-eqn}) to find a change of variables $\eta = h(X,\tau)$,
$\xi = \tau$. Taking $\beta_{0} = c = 0$, one obtains
\[
\eta = X - \ln \left| 1 - \tau e^{X - \lambda T} \right| \, , \; \; \; \; \; \; \xi = \tau \; ,
\]
and substituting into (\ref{new-gamma})--(\ref{new-beta}) one finds the (pseudo)potentials 
$\gamma$, $\delta$ and $\beta$ as functions of $X$, $T$, and $\tau$:
\begin{equation}
\gamma (X, T,\tau) = \lambda \; \; \; \; \; \; 
\delta (X, T, \tau) = \lambda (X - \lambda T) \; \; \; \; \; 
\beta (X, T, \tau) = 0 \; .
\end{equation}
Corollary \ref{newsol} then implies that a two--parameters family of solutions of the 
Camassa-Holm equation
\[
m = U_{XX} - U \; , \; \; \; \; \; \; \; \; m_{T} = - m_{X} U - 2 m U_{X} \; ,
\] 
is determined by the initial value problem
\[
\frac{\partial \, U}{\partial \tau} =
                 \gamma (X,T,\tau)\,{\rm e}^{\displaystyle (1/\lambda)\,\delta(X,T,\tau)} \; ,
 \; \; \; \; \;
U(X,T,0) = {\rm e}^{X},
\]
since at $\tau = 0$ the independent variables $X$ and $\eta$ coincide. One finds
\[
u(X,T,\tau) = \lambda \tau e^{X - \lambda T} + e^{X}.
\]
\end{example}

\begin{example}
Consider the nonlinear equation (\ref{gen-sys}),
\begin{equation}
m = \nu^{2}U_{\eta\eta} - \epsilon U \, , \; \; \; \; \; \; \; \; 
m_{T} = - m_{\eta}U - 2m U_{\eta}+\frac{2}{3}(1-\nu^{5/2})U_{\eta\eta\eta}\; , \label{gen-sys-ex}
\end{equation}
and the trivial solution $U(\eta,T) = 0$. The corresponding (pseudo)potentials $\gamma_{0}$, 
$\delta_{0}$ and $\beta_{0}$  which one obtains from (\ref{g-x})--(\ref{b-t}) can be chosen to be
\begin{eqnarray}
\gamma_{0} & = & \frac{2}{3}\,\sqrt {1+\zeta\,{\nu}^{2}}\sqrt {\epsilon\,\left (\zeta+\sqrt {
\nu}\right )} \; , \\
\delta_{0} & = & \frac{2}{3}\,\sqrt {1+\zeta\,{\nu}^{2}}\sqrt {\epsilon}\sqrt {\zeta+\sqrt {\nu
}}\left (\eta-\frac{2}{3}\,\left (\zeta+\sqrt {\nu}\right )t\right ) \; , \\
\beta_{0} & = & \frac{1}{3}\,\left ({\nu}^{5/2}-1\right )\sqrt {\epsilon}
\frac{\sqrt {1+\zeta\,{\nu}^{2}}}{\sqrt {\zeta+\sqrt {\nu}}}\,
\exp\left({\frac {\left (\eta-(2/3)\,\left (\zeta+\sqrt {\nu}\right )t\right)
\sqrt {\epsilon}\sqrt {\zeta+\sqrt {\nu}}}{\sqrt {1+\zeta\,{\nu}^{2}}}}\right) \; .
\end{eqnarray}
Proposition \ref{sy2} then yields expressions for the (pseudo)potentials $\gamma$, 
$\delta$ and $\beta$ as functions of $\eta$, $\xi$ and $T$, and Equation (\ref{tra}) and
Corollary \ref{newsol} allow one to find a two--parameters family of solutions to the system
(\ref{gen-sys-ex}). 

Consider, for instance, the $\nu = 0$ case. Equation (\ref{tra}) implies that in this 
case the ``old'' and ``new'' independent variables $\eta$ and $X$ agree. One then finds 
$\gamma(X,T,\tau)$, $\delta(X,T,\tau)$ and $\beta(X,T,\tau)$ to be
\begin{eqnarray}
\gamma & = & -\frac{2}{3}\,{\frac {\sqrt {\zeta} 
    \left (\xi\,{{\rm e}^{-(1/3)\,\sqrt {\zeta}\left(-3\,X+2\,T\zeta\right )}}
   -4\,\sqrt {\zeta}\right )}{\xi\,{{\rm e}^{-(1/3)\,\sqrt {\zeta}\left(-3\,X+2\,T\zeta\right )}}
       +4\,\sqrt {\zeta}}}              \; ,    \\
\delta & = &  \frac{8}{3}\,\ln (2)+\frac{2}{3}\,\ln \left|{\frac {{{\rm e}^{-(1/3)\,\sqrt {\zeta}\left (-3\,X+
2\,T\zeta\right )}}\zeta}{\left (\xi\,{{\rm e}^{-(1/3)\,\sqrt {\zeta}\left (-3
\,X+2\,T\zeta\right )}}+4\,\sqrt {\zeta}\right )^{2}}} \right|    \; ,                \\
\beta & = &  -\frac{4}{3}\,{\frac {{{\rm e}^{-(1/3)\,\sqrt {\zeta}\left (-3\,X+2\,T\zeta\right )
}}}{\xi\,{{\rm e}^{-(1/3)\,\sqrt {\zeta}\left (-3\,X+2\,T\zeta\right )}}+4\,\sqrt {\zeta}}} \; ,
\end{eqnarray}
and it follows from Corollary \ref{newsol} that the function
\begin{equation}
U(X,T,\tau) = 
{\frac {32}{3}}\,{\frac {{\zeta}^{3/2}{{\rm e}^{ -(1/3)\,\sqrt {\zeta}\left (-3
\,X+2\,T\zeta\right )}}\xi}{\left (\xi\,{{\rm e}^{-(1/3)\,\sqrt {\zeta}
\left (-3\,X+2\,T\zeta\right )}}+4\,\sqrt {\zeta}\right )^{2}}} \; , 
\end{equation}
solves the KdV equation 
\[
3\,\left ({\frac {\partial }{\partial X}}U(X,T)\right )U(X,T)
+ \frac{2}{3}\,{\frac {\partial ^{3}}{\partial {X}^{3}}}U(X,T)+{\frac {
\partial }{\partial T}}U(X,T)=0 \; .
\]
This is a travelling wave solution if $\xi > 0$, and a 
singular solution for some negative values of the parameter $\xi$.
\end{example}

It is of course of interest to investigate whether the important peakon solutions of the 
Camassa--Holm and Hunter--Saxton equations \cite{CH,bss,bss1}, appear within the approach
considered in this work. This problem will be treated in a separated publication.

\subsection*{Acknowledgements}
The author thanks R. Beals, G. Misio{\l}ek, and J. Szmigielski for interesting questions
and discussions on the equations considered in this paper, and for sharing with him their
preprints \cite{bss1} and \cite{KM}.

\end{document}